\documentclass[lettersize,journal]{IEEEtran}
\usepackage{graphicx}                                           
\usepackage{subfigure}
\usepackage{booktabs}                                           
\usepackage{makecell}                                           
\usepackage{colortbl}                                           
\usepackage{multirow}                                           
\usepackage{cite}
\usepackage{amsmath}
\usepackage{amssymb}
\usepackage{amsthm}
\usepackage{mathtools}
\usepackage{csquotes}
\usepackage{threeparttable}
\usepackage{subcaption}
\usepackage{xpatch}
\usepackage[inline]{enumitem}
\usepackage{url}
\usepackage{tabularx}
\usepackage{todonotes}
\usepackage{balance}

\newtheorem{definition}{\bf Definition}

\usepackage{tikz}
\usetikzlibrary{arrows.meta}
\usetikzlibrary{trees, positioning}

\AtBeginDocument{%
  }

\newcommand{\tabincell}[2]{\begin{tabular}{@{}#1@{}}#2\end{tabular}}

\begin{document}

\title{A Survey of Densest Subgraph Discovery on Large Graphs}



\author{
\IEEEauthorblockN{
    Wensheng Luo\textsuperscript{1},
    Chenhao Ma\textsuperscript{2},
    Yixiang Fang\textsuperscript{2},
    and Laks V. S. Lakshmanan\textsuperscript{3}
}\\
\IEEEauthorblockA{\textsuperscript{1}Hunan University};
\IEEEauthorblockA{\textsuperscript{2}The Chinese University of Hong Kong, Shenzhen};
\IEEEauthorblockA{\textsuperscript{3}University of British Columbia}\\
\IEEEauthorblockA{luowensheng@hnu.edu.cn; machenhao@cuhk.edu.cn; fangyixiang@cuhk.edu.cn; laks@cs.ubc.ca}
}


\maketitle

    \begin{abstract}

With the prevalence of graphs for modeling complex relationships among objects, graph mining has attracted a great deal of attention from academic and industrial communities in recent years. As one of the most fundamental problems in graph mining, the {\it densest subgraph discovery} (DSD) problem has found a wide spectrum of real applications, such as the discovery of filter bubbles in social media, finding groups of actors propagating misinformation in social media, social network community detection, graph index construction, regulatory motif discovery in DNA, fake follower detection, and so on. Theoretically, DSD closely relates to other fundamental graph problems, such as network flow and bipartite matching. Triggered by these applications and connections, DSD has garnered much attention from the database, data mining, theory, and network communities.

In this survey, we first highlight the importance of DSD in various real-world applications and the unique challenges that need to be addressed. Subsequently, we classify existing DSD solutions into several groups, which cover around 50 research papers published in many well-known venues (e.g., SIGMOD, PVLDB, ICDE, TODS, WWW), and conduct a thorough review of these solutions in each group. Afterwards, we analyze and compare the models and solutions in these works. Finally, we point out a list of promising future research directions.
It is our hope that this survey not only helps researchers have a better understanding of existing densest subgraph models and solutions but also provides insights and identifies directions for future study.

\end{abstract}


    \section{Introduction}
\label{sec:intro}

\begin{figure}[h]
\centering
    \subfigure[An undirected graph]{
        \centering
    	\includegraphics[width=0.3\linewidth]{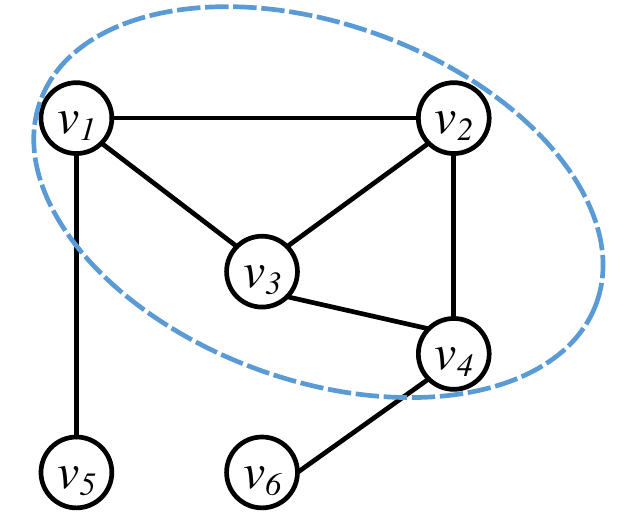}
    	\label{fig:un-graph}
    }
    \subfigure[A directed graph]{
        \centering
    	\includegraphics[width=0.3\linewidth]{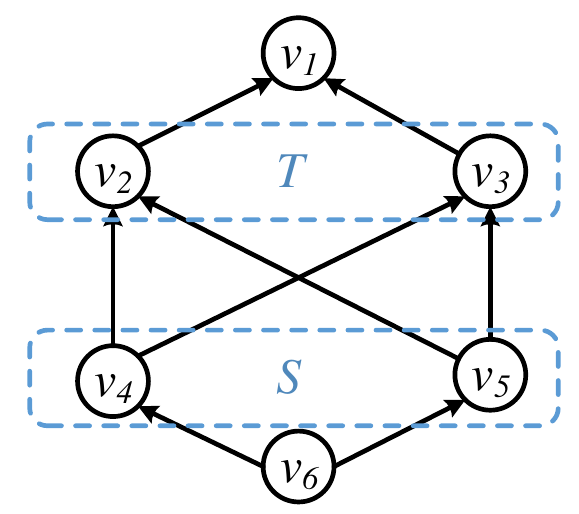}
    	\label{fig:di-graph}
    }
\caption{Examples of undirected and directed graphs.}
\label{fig:graphs}
\end{figure}

In emerging systems that manage complex relationships among objects, different kinds of graphs are often used to model relationships between objects \cite{ching2015one,java2007we,karlebach2008modelling,ma2019linc,albert1999diameter}. For example, the Facebook friendship network can be modeled as an undirected graph by mapping users to vertices and friendships to edges \cite{ching2015one}. Fig. \ref{fig:un-graph} illustrates an undirected graph of friendship, where $v_1$ and $v_2$ have an edge meaning that they are friends. In Twitter, a directed edge can represent the ``following'' relationship between two users \cite{java2007we}. Fig. \ref{fig:di-graph} gives an example of a directed graph. In gene regulatory networks, a link from gene A to gene B denotes the regulatory relationship between those genes \cite{karlebach2008modelling}. Moreover, the Web network can also be modeled as a vast directed graph \cite{albert1999diameter}.

\begin{table*}[t]
\centering
\setlength{\tabcolsep}{2.5pt}
\renewcommand{\arraystretch}{1.05}
\caption{{\color{black}Classification of existing DSD works on undirected graphs.
The ``original DSD problem" means that given an undirected graph, return the subgraph with the largest edge-density. $n$ is the number of vertices; $k$ is an integer; $\epsilon>0$ is a real value; the approximation ratio is defined as the density ratio between the returned subgraph and the optimal DS.}}
\label{tab:models_undirected}

\begin{tabular}{c|c|c}
\hline
\multicolumn{2}{c|}{\bf Original DSD problem} &
\multirow{2}{*}{\tabincell{c}{\bf Variants of original\\ \bf DSD problem}} \\
\cline{1-2}
{\bf Exact solutions} & {\bf Approx. solutions} & \\
\hline\hline

\tabincell{c}{
Unweighted case~\cite{goldberg1984finding,charikar2000greedy,fang2019efficient}\\
Non-negative weighted case~\cite{danisch2017large}
}
&
\tabincell{c}{
2-approximation~\cite{charikar2000greedy,fang2019efficient,boob2020flowless,luo2023scalable}\\
2(1+$\epsilon$)-approximation~\cite{bahmani2012densest,tsourakakis2019novel}\\
$(1+\epsilon)$-approximation \cite{sawlani2020near,boob2020flowless,chekuri2022densest,harb2022faster,xu2023efficient}\\
DS maintenance~\cite{bahmani2012densest,epasto2015efficient,bhattacharya2015space,hu2017maintaining,sawlani2020near,esfandiari2015applications,mcgregor2015densest,chekuri20221}
}
&
\tabincell{c}{
Clique-density-based DSD~\cite{tsourakakis2015k,mitzenmacher2015scalable,samusevich2016local,fang2019efficient,sawlani2020near,sun2020kclist++,zhou2024counting}\\
Pattern-density-based DSD~\cite{fang2019efficient}\\
Densest $k$-subgraph~\cite{asahiro2000greedily,bhaskara2010detecting,bhaskara2012polynomial,bourgeois2013exact,nonner2016ptas,kawase2018densest,gonzales2019densest,bonchi2021finding}\\
Size-bounded DSD~\cite{andersen2009finding}\\
Top-$k$ overlapping DSD~\cite{galbrun2016top,dondi2021novel,dondi2021top}\\
Maximum total density DSD~\cite{balalau2015finding}\\
Density-friendly graph decomposition~\cite{tatti2015density,danisch2017large}\\
Locally DSD~\cite{qin2015locally,ma2022finding}\\
DS deconstruction~\cite{chang2020deconstruct}\\
Top-$k$ DSD maintenance~\cite{nasir2017fully}\\
Anchored densest subgraph search~\cite{dai2022anchored,ye2024efficient}\\
Differential privacy DSD~\cite{nguyen2021differentially,dhulipala2022differential}\\
Densest diverse subgraphs search~\cite{anagnostopoulos2020spectral,miyauchi2023densest}
}
\\
\hline
\end{tabular}
\end{table*}

\begin{table}[t]
\centering
\setlength{\tabcolsep}{2pt}
\renewcommand{\arraystretch}{1}
\caption{{\color{black}Classification of existing DSD works on directed graphs. $k_1$ and $k_2$ are integers.}}
\label{tab:models_directed}
\resizebox{\columnwidth}{!}{
\begin{tabular}{c|c|c}
\hline
\multicolumn{2}{c|}{\bf Original DSD problem} & \multirow{2}{*}{\bf Variants} \\
\cline{1-2}
{\bf Exact solutions} & {\bf Approx. solutions} & \\
\hline\hline

\tabincell{c}{
Unweighted \\ case~\cite{kannan1999analyzing,charikar2000greedy,khuller2009finding,ma2020efficient,ma2021directed,luo2023scalable}\\
Non-negative \\ weighted case~\cite{ma2021directed}
}
&
\tabincell{c}{
$O(\log n)$-approximation~\cite{kannan1999analyzing}\\
2-approximation~\cite{charikar2000greedy,ma2020efficient,ma2021directed}\\
2(1+$\epsilon$)-approximation~\cite{bahmani2012densest}\\
$(1+\epsilon)$-approximation~\cite{sawlani2020near,ma2022convex}\\
DS maintenance~\cite{bahmani2012densest,sawlani2020near,ma2021directed}
}
&
\tabincell{c}{
Densest at least \\ $k_{1}$, $k_{2}$ subgraph~\cite{khuller2009finding}
}
\\
\hline
\end{tabular}
}
\end{table}

\begin{table}[t]
\centering
\setlength{\tabcolsep}{2pt}
\renewcommand{\arraystretch}{1}

\caption{{\color{black}Classification of existing DSD works on other graph types.}}
\label{tab:models_others}

\resizebox{\columnwidth}{!}{
\begin{tabular}{c|c|c}
\hline
\multicolumn{2}{c|}{\bf Original DSD problem} & \multirow{2}{*}{\bf Variants}\\
\cline{1-2}
{\bf Exact solutions} & {\bf Approx. solutions} \\
\hline\hline

\tabincell{c}{
Uncertain graphs~\cite{zou2013polynomial}\\
HINs~\cite{chen2023densest}\\
Hypergraphs~\cite{hu2017maintaining,HuangGV24}
}
&
\tabincell{c}{
Bipartite graphs~\cite{andersen2010local,mitzenmacher2015scalable,hooi2016fraudar}\\
Multilayer graphs~\cite{jethava2015finding,galimberti2017core,galimberti2020core}\\
Uncertain graphs~\cite{miyauchi2018robust}\\
HINs~\cite{chen2023densest}\\
Hypergraphs~\cite{hu2017maintaining,arafat2023neighborhood,bera2022new,HuangGV24}
}
&
\tabincell{c}{
Dense connected \\ subgraphs~\cite{wu2015finding} \\
Anchored hyper DS \cite{HuangGV24}
}
\\
\hline

\end{tabular}
}
\end{table}

As one of the most fundamental problems in graph data mining, the {\it Densest Subgraph Discovery} (DSD) problem aims to discover a very ``dense'' subgraph from a given graph. More precisely, given an undirected graph, the DSD problem \cite{goldberg1984finding} finds a subgraph with the highest {\em edge-density}, which is defined as the number of edges over the number of vertices in the subgraph, and it is often termed as the densest subgraph (DS).
This problem has also been extensively studied on other kinds of graphs, including directed, uncertain, bipartite, and multi-layer graphs.
To provide a structured overview, we summarize representative DSD works in Tables~\ref{tab:models_undirected}--\ref{tab:models_others}, covering undirected graphs, directed graphs, and other graph types, and organizing methods by problem formulations and solution paradigms.

Note that DSD is closely related to a broader class of cohesive subgraph models. 
Various approaches, such as $k$-core, $k$-truss, and $k$-ECC, have been proposed to characterize dense regions in graphs~\cite{fang2020survey,changcohesive}. 
These models rely on local structural constraints, whereas the densest subgraph model is defined by a global density objective based on the edge-to-vertex ratio. 
Accordingly, different from the previous survey on community
search~\cite{fang2020survey}, we focus specifically on
density-based optimization problems, their algorithmic paradigms and theoretical guarantees, and their extensions across different graph models in this paper.

The DSD problem lies in the core of graph mining \cite{bahmani2012densest,gionis2015dense}, and is widely used in network science \cite{chen2010dense,tsourakakis2013denser,hooi2016fraudar,angel2014dense}, graph databases \cite{jin20093,cohen2003reachability,zhang2012extracting,zhao2012large,buehrer2008scalable}, biological analysis \cite{fratkin2006motifcut,saha2010dense}, 
information dissemination analysis to discover filter bubbles and groups of actors propagating misinformation 
\cite{lakshmanan2022quest}, and system optimization \cite{gionis2013piggybacking,gionis2015dense,gibson2005discovering}.
Here is a list of typical applications, to name a few:

\begin{itemize}
    \item \textbf{Network analysis.} In social networks (e.g., Facebook), the DS discovered can be used to find the ``closely connected groups'', since these groups correspond to network communities~\cite{chen2010dense,tsourakakis2013denser}. Besides, the DS has proven effective for detecting network anomalies, such as revealing fake followers in follower/followee networks \cite{ma2020efficient} and detecting fake accounts in e-commerce networks~\cite{beutel2013copycatch}.
    
    \item \textbf{Graph databases.} Solution to the  DSD problem is a building block for solving many graph problems, such as reachability queries~\cite{cohen2003reachability} and motif detection~\cite{fratkin2006motifcut,saha2010dense}. For example, the 2-hop-cover-based index is an efficient index to answer whether a target node $t$ is reachable from a source node $s$. However, finding a minimum 2-hop cover of a set of shortest paths is NP-hard, and the DS can be used to find an approximation solution with ratio $O(\log n)$~\cite{cohen2003reachability}.

    \item \textbf{Biological data analysis.} 
    %
    {\color{black}
    DSD solutions have been shown useful for identifying regulatory motifs in genomic DNA \cite{fratkin2006motifcut}, and gene annotation graphs \cite{saha2010dense}. 
    For example, Fratkin et al. \cite{fratkin2006motifcut} proposed the MotifCut system, which converts DNA sequences into a set of $k$-mers and constructs a graph where each $k$-mer corresponds to a vertex and two $k$-mers are linked if their nucleotide similarity is high. 
    Regulatory motif discovery is then formulated as a densest subgraph search solved via DSD algorithms. 
    On yeast regulatory sequences with tens of thousands of vertices, the method achieved higher motif discovery accuracy than established tools while maintaining comparable runtime.
    In addition, MotifCut yields motifs that differ from those of traditional methods, likely due to restrictive assumptions such as positional independence, which can miss complex dependencies, whereas the DSD formulation better captures such patterns.
    }

    \item \textbf{Filter bubbles and misinformation.} 
    Social networks allow users to share news/views with many peers, but they are also known to contribute to and exacerbate the problem of filter bubbles and echo chambers, which tend to reinforce preexisting opinions and beliefs and spread misinformation. 
    DSs in social networks have proven useful for identifying echo chambers and groups of actors engaged in spreading misinformation~\cite{memon2020characterizing,lakshmanan2022quest,fazzone2022discovering}.
    {\color{black}
    For example, the DSAR model proposed in \cite{fazzone2022discovering} formulates the problem as a density optimization task that simultaneously considers proximity to attractor nodes and distance from repulser nodes. 
    Experiments on 26 real-world networks, including graphs with up to 276,657 vertices and 2.91 billion edges, show that the proposed algorithm achieves near-optimal solutions.
    }
\end{itemize}

Besides, the DSD problem is also closely related to other fundamental graph problems, such as network flow and bipartite matching \cite{sawlani2020near}.
Due to the theoretical and practical importance, researchers from the database, data mining, computer science theory, and network communities designed efficient and effective solutions to the DSD problem.

Despite the high importance of DSD, the DSD problem is very challenging: 
Firstly, the exact DSD solutions (e.g., \cite{goldberg1984finding,fang2019efficient}) often involve the computation of maximum network flow which has a very high time complexity, while many real-world graphs are often with huge sizes (e.g., Facebook has more than 2.89 billion monthly active users as of October 2021\footnote{\url{https://www.statista.com/statistics/272014/global-social-networks-ranked-by-number-of-users/}}).
Thus, the first key challenge is how to develop efficient algorithms. Many researchers have developed various techniques as shown in the literature~\cite{fang2019efficient,ma2020efficient,charikar2000greedy,bahmani2012densest}.
Secondly, many real-world networks do not simply fall into one of the categories -- undirected or directed graphs. Furthermore,  a real application often needs not just one single DS, but typically the top-$k$ DSs. For instance, to discover echo chambers in social networks, one often needs to explore the top-$k$ DS for some $k$ and analyze them further. A similar comment applies to the application of community detection. On the other hand, existing solutions to the original DSD problem studied on undirected and directed graphs are only able to return one single DS.
Therefore, the second challenge is how to perform effective DSD such that it can well satisfy the specific requirements on different graphs.
To this end, some researchers have extended the original DSD problem formulation and solutions for bipartite graphs (e.g., \cite{andersen2010local}), multilayer graphs (e.g., \cite{jethava2015finding, galimberti2020core, DBLP:journals/pvldb/BehrouzHL22,DBLP:conf/www/HashemiBL22}), and uncertain graphs (e.g., \cite{DBLP:conf/kdd/BonchiGKV14, DBLP:conf/sigmod/HuangLL16, miyauchi2018robust}).
In addition, many variants of the DSD have been studied to satisfy different practical requirements~\cite{tsourakakis2015k,fang2019efficient,asahiro2000greedily,tatti2015density,qin2015locally,chang2020deconstruct}.

In summary, many existing works have studied the DSD problem extensively from different aspects, and there is a lack of a systematic review and a comparative study among them, except for a few preliminary works \cite{gionis2015dense,farago2019search,lanciano2023survey,zhou2024depth} which are different from ours, as we will analyze later in Section \ref{sec:related-DSD}.
To this end, in this paper, we aim to provide a comprehensive review of works on  {\em densest} subgraph discovery, which directly use the {\em edge-density} definition, or density definitions extended from it. 
{\color{black}
Our review covers representative results published up to January 2025. Works appearing after this cut-off are beyond the scope of the present survey.
}
There are many works on related concepts of $k$-core, $k$-truss, $k$-clique, $k$-ECC, etc~\cite{fang2020survey}. Given the volume of the body of work on DSD based on edge-density, such dense subgraph models will not be discussed in detail in this paper.
An earlier version of this paper has been published in a tutorial of a previous conference~\cite{FangLM22}, which serves as a foundation for the present study.
{\color{black}
We refer readers to a recent large-scale empirical study \cite{zhou2024depth}, which provides a systematic experimental comparison of representative algorithms for the original UDS and DDS problems under a unified framework.
}

The principal contributions of the paper are as follows:

{\color{black}
\begin{itemize}
    \item \textbf{A principled taxonomy of DSD formulations and solution paradigms.} We establish a structured taxonomy that organizes over 50 representative studies into two fundamental categories: the original DSD formulation and its variant extensions. Distinct from prior surveys, our framework explicitly disentangles problem definitions from algorithmic paradigms and systematically categorizes them across undirected, directed, and other graph settings. This taxonomy provides a coherent structural view of the DSD landscape and clarifies the relationships among heterogeneous formulations.

    \item \textbf{A systematic cross-model comparison of theoretical and algorithmic properties.} We systematically analyze and compare the efficiency and effectiveness of different DSD formulations across multiple analytical dimensions, including density definitions, algorithmic time complexities, and theoretical approximation guarantees. This comparative synthesis highlights the trade-offs among computational cost, solution quality, and scalability across different DSD formulations.
    
    \item \textbf{A consolidated research perspective and reproducibility support.} We synthesize unresolved theoretical challenges, scalability limitations, and emerging research directions, including dynamic, distributed, and application-driven extensions of DSD. To enhance reproducibility and lower the barrier for experimental evaluation, we further curate publicly accessible implementations of representative algorithms and consolidate them into an openly available GitHub repository\footnote{https://github.com/GearlessL/DSD-algorithm-collection/} as a centralized reference resource.
\end{itemize}
}


    \section{Problem statements}
\label{sec:problem}

In this section, we formally present the original definitions of graph density and DSD problems on both undirected graphs and directed graphs.

\subsection{Problem statements}

\begin{definition}[\textbf{Edge-density on undirected graphs }~\cite{goldberg1984finding}]
	\label{def:edge-udensity}
	Given an undirected graph $G$=$(V,E)$, its edge-density $\rho(G)$ is defined as the number of edges over the number of vertices
\begin{equation}
     \rho(G)= \frac{{|E|}}{{|V|}}.
     \label{eq:density}
\end{equation}
\end{definition}

\begin{definition}[\textbf{Undirected Densest Subgraph (UDS) problem}~\cite{goldberg1984finding,tsourakakis2015k,fang2019efficient}]
\label{def:densest}
Given an undirected graph, find the subgraph whose corresponding edge-density is the highest among all the possible subgraphs, also called the undirected densest subgraph (UDS).
\end{definition}

For example, in the undirected graph of Fig.~\ref{fig:un-graph}, the density of the subgraph in the dashed ellipse is 5/4, since there are five edges and four vertices, and it is the densest subgraph (DS) because its density is the highest among all possible subgraphs.

The density of a directed graph is defined over two vertex sets with the concept of ($S$, $T$)-induced subgraph. 
Given a directed graph $G$=$(V,E)$ and two vertex sets $S$ and $T$, an ($S$, $T$)-induced subgraph, denoted by $G[S,T]$, is a subgraph consisting of two vertex sets $S, T \subseteq V$ and an edge set $E(S, T) {=} E \cap (S \times T)$.

\begin{definition}[\textbf{Edge-density on directed graphs}~\cite{kannan1999analyzing,khuller2009finding,ma2020efficient,ma2021directed}]
\label{def:edge-ddensity}
Given a directed graph $G$=$(V,E)$ and two vertex sets $S$ and $T$, the edge-density of an ($S$, $T$)-induced subgraph $\rho(S, T)$ is the number of edges linking vertices in $S$ to the vertices in $T$ over the square root of the product of their sizes
\begin{equation}
\rho(S, T)= \frac{{|E(S, T)|}}{{\sqrt{|S||T|}}}.
\label{eq:didensity}
\end{equation}
\end{definition}

\begin{definition}[\textbf{Directed Densest subgraph (DDS) problem }\cite{kannan1999analyzing,gionis2015dense,charikar2000greedy,khuller2009finding,bahmani2012densest}]
\label{def:densest}
Given a directed graph $G$, find the subgraph whose corresponding edge-density is the highest among all the possible subgraphs, also called the directed densest subgraph (DDS).
\end{definition}

For instance, in the directed graph of Fig. \ref{fig:di-graph}, for the two vertex sets $S$=$\{v_4,v_5\}$ and $T$=$\{v_2,v_3\}$, the density of the ($S$, $T$)-induced subgraph is $\rho(S,T)$ = $\frac{4}{\sqrt{2\times 2}}$ = 2, since there are four edges linking from $S$ to $T$. This subgraph is the DS because there are no other two vertex sets having a higher density.

For ease of exposition, we will use $G[S^{*}]$ and $G[S^{*}, T^{*}]$ to denote the DSs in the undirected and directed graphs, respectively.

{\color{black}
\subsection{Flow-based Density Feasibility Paradigm}
\label{subsec:flow-prelim}
A key idea in densest subgraph algorithms is to reduce density maximization to a sequence of \emph{feasibility tests}: given a value $g$, determine whether a subgraph with density at least $g$ exists.
For undirected graphs, Goldberg~\cite{goldberg1984finding} reduces this test to a minimum $s$--$t$ cut problem. 
Given $G=(V,E)$ with $|E|=m$, construct a flow network by adding a source $s$ and sink $t$. Each vertex $v$ connects to $s$ with capacity $m$ and to $t$ with capacity $m+2g-d_v$, where $d_v$ is the degree of $v$. Each undirected edge is replaced by two unit-capacity arcs. 
A subgraph of density at least $g$ exists iff the minimum cut has capacity less than $m|V|$ (Fig.~\ref{unflow}). This reduction underlies many exact and approximation algorithms.

\begin{figure}[h]
	\centering
	\includegraphics[width=.6\linewidth]{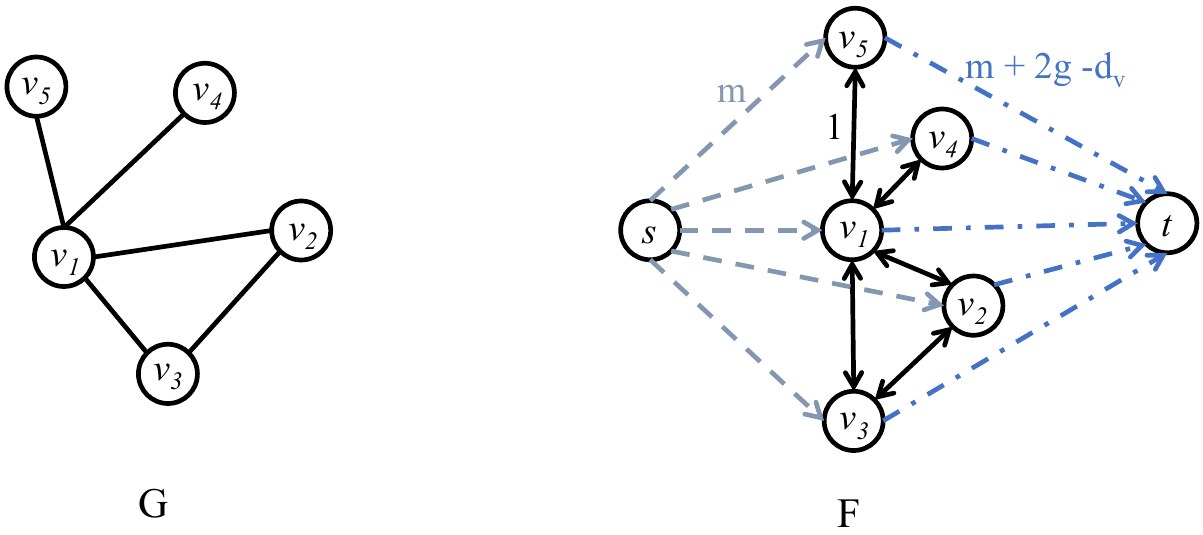}
	\caption{An undirected Graph $G$ and its flow network $F$.}
	\label{unflow}
\end{figure}

For directed graphs, the same paradigm applies but must encode directionality and the interaction between two vertex sets. This is achieved by duplicating each vertex into two layers. 
As shown in Fig.~\ref{fig:di-flow-net}, vertices are split into sets $A$ and $B$. Each original edge induces an arc from $B$ to $A$ with capacity 2. Edges from $s$ to $A$ have capacity $m$, while edges from $B$ to $t$ depend on parameters $g$ and $a$.

\begin{figure} [ht]
    \centering
    \subfigure[Directed graph]{\includegraphics[width=0.24\linewidth]{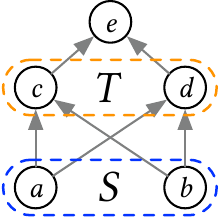}}
    \subfigure[Flow network]{\includegraphics[width=0.58\linewidth]{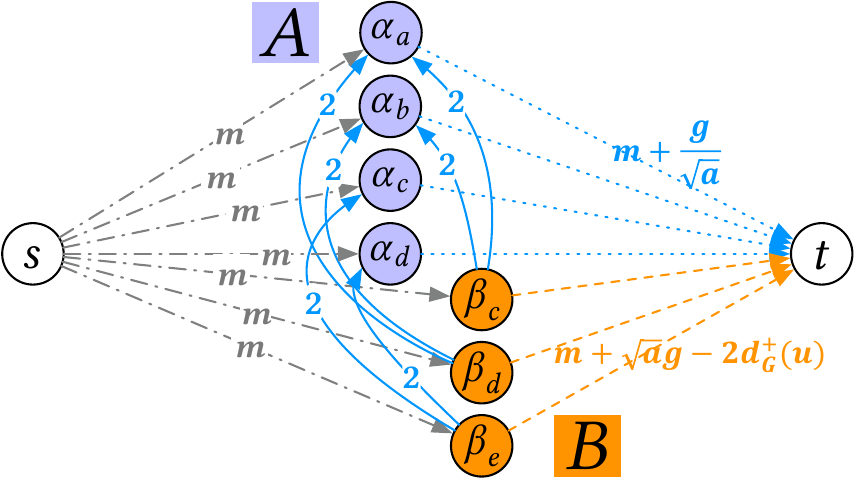}}
    \caption{Flow network built for the directed graph}
    \label{fig:di-flow-net}
\end{figure}


This paradigm extends to higher-order density notions (e.g., clique density) by constructing flow networks on instance graphs, where nodes represent higher-order structures. 
For $k$-clique density, $(k{-}1)$-cliques are treated as nodes. For each vertex $v$, add edges $(S,v)$ with capacity $\deg(v,\Psi)$ and $(v,T)$ with capacity $g|V_{\Psi}|$. 
For each $(k{-}1)$-clique $\psi_i$, add infinite-capacity edges $\psi_i \to v$ if $v \in \psi_i$, and unit-capacity edges $v \to \psi_i$ if $v \cup \psi_i$ forms a $k$-clique, as shown in Fig.~\ref{fig:cflow}.

\begin{figure}[h]
  \centering
  \subfigure[graph]{\includegraphics[width=0.1\textwidth]{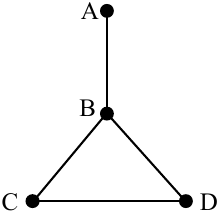}}
  \quad
  \subfigure[2-clique]{\includegraphics[width=0.1\textwidth]{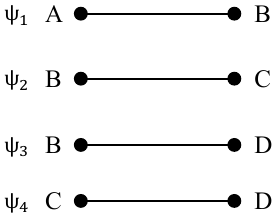}}
  \quad
  \subfigure[the flow network]{\includegraphics[width=0.2\textwidth]{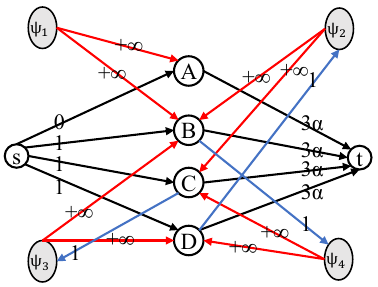}}
  \caption{An undirected Graph $G$ and its flow network where $\Psi$ is a triangle.}
  \label{fig:cflow}
\end{figure}

{\bf Max-flow solvers.}
The efficiency of flow-based densest subgraph algorithms depends critically on the performance of the underlying maximum flow or minimum cut routines. 
In this paper, we use $O(t_{\text{Flow}})$ to denote the time cost of a max-flow computation.
In practice, classical combinatorial algorithms such as Dinic’s algorithm \cite{dinic1970algorithm} ($O(n^2m)$) and push-relabel \cite{goldberg1988new} ($O(n^3)$) are widely adopted due to their simplicity, robustness, moderate memory usage, and strong empirical performance.

{\bf Remark.}
Recent theoretical advances show that the exact maximum flow problem can be solved in $O(mn)$ time \cite{orlin2013max}, and further improved to almost-linear time $O(m^{1+o(1)})$ by recent works~\cite{chen2022maximum}.
However, such algorithms rely on highly sophisticated frameworks and may incur large constant factors, substantial memory overhead, and significant implementation complexity. In particular, these methods are often difficult to implement and are primarily of theoretical interest rather than practical use. In practice, classical algorithms (e.g., push–relabel methods such as Goldberg–Tarjan) often remain more efficient except on extremely large-scale instances.
Therefore, for densest subgraph problems, practical performance should not be inferred solely from asymptotic bounds, and careful consideration of implementation efficiency is necessary.

\begin{table}[h]
\centering
\caption{{\color{black}Frequently used notations and their meanings.}}
\label{tab:notations}
\begin{tabular}{l | p{0.76\columnwidth}}
\hline
\textbf{Notations} & \textbf{Meaning} \\
\hline
\hline
$G=(V,E)$ & Input graph with vertex set $V$ and edge set $E$ \\
\hline
$n, m$ & Number of vertices and edges in $G$ \\
\hline
$\Delta(G)$ & Maximum degree of graph $G$ \\
\hline
$\alpha(G)$ & The arboricity of an undirected graph $G$, i.e., minimum number of forests into which its edges can be partitioned \\
\hline
$c_k$ & The number of $k$-clique instances in $G$ \\
\hline
$\rho(G)$ & Edge-density of graph $G$, defined as $|E|/|V|$ \\
\hline
$E(S,T)$ & Set of edges from $S$ to $T$ for a directed graph \\
\hline
$\rho(S,T)$ & Density of a directed $(S,T)$-induced subgraph \\
\hline
$\rho^*$ & Optimal density of the densest subgraph \\
\hline
$d_v$ & Degree of vertex $v$ \\
\hline
$k_{\max}$ & Maximum core number in the graph \\
\hline
$\epsilon$ & Approximation parameter controlling accuracy \\
\hline
$t_{\text{Flow}}$ & Time complexity of the max-flow (or min-cut) computation \\
\hline
\end{tabular}
\end{table}
}



    \section{DSD on undirected graphs}
\label{sec:outlineUndirected}

In this section, we mainly review the works that study the original UDS problem and its variants on undirected graphs.
%
We classify the solutions to the original UDS problem as exact solutions, approximation solutions, and maintenance solutions.

\subsection{Exact solutions}


Exact solutions to the UDS problem can be divided into two main categories: 1) flow network-based solutions~\cite{goldberg1984finding,gallo1989fast,fang2019efficient} and 2) linear programming-based solutions~\cite{charikar2000greedy}.
In these solutions, Goldberg~\cite{goldberg1984finding} and Fang et al.~\cite{fang2019efficient} proposed algorithms with a worst-case running time of $O(\log n \cdot t_{\text{Flow}})$, while Gallo et al.~\cite{gallo1989fast} introduced a parametric max-flow algorithm with time complexity of $O(\alpha \cdot t_{\text{Flow}})$, where $\alpha$ is a constant.
Here, $t_{\text{Flow}}$ represents the time cost of a max-flow algorithm.
Notably, Gallo’s algorithm achieves the best theoretical time cost, whereas Fang et al.'s algorithm demonstrates superior practical performance.
We now sequentially review the exact solutions to the UDS problem.

(1) Goldberg~\cite{goldberg1984finding} first introduced the edge-density and then formally defined the UDS problem. The author proposed an exact solution, namely \texttt{Exact}, based on maximum flow, consisting of three key steps:

\begin{itemize}
    \item[a)] Guess the density $g$ of the DS through binary search, where $g \in [0, d_m]$ and $d_m$ is the maximum degree of vertices in $V$;
    \item[b)] Build a flow network based on the undirected graph and guessed maximum density $g$;
    \item[c)] Verify whether $g$ is the maximum edge-density value by computing the max-flow (min-cut) of the flow network. 
\end{itemize}

%
%
In step b), the flow network is constructed from the original graph, as illustrated in Fig.~\ref{unflow}.
After constructing the flow network, \texttt{Exact} verifies whether $g$ is maximum by computing the minimum cut of the flow network.
The binary search on $g$ needs to be performed at most $O(\log n)$ times until the gap between the upper and lower bounds of $g$ is less than $\frac{1}{n(n-1)}$.
Thus, the time complexity of \texttt{Exact} is $O(t_{\text{Flow}} \cdot\log n)$, where $t_{\text{Flow}}$ is the time cost of computing the min-cut of a flow network.
%
%

Gallo et al.~\cite{gallo1989fast} introduced a parametric max-flow algorithm, which can also find the DS since it is a special case of the fractional programming problem, and achieves higher efficiency.
The authors proposed a parametric max-flow algorithm that avoids the need to reconstruct bipartite networks or solve minimum-cut problems.
%
%
This approach achieves a time complexity of $O(\alpha \cdot t_{\text{Flow}})$, where $\alpha$ is a constant, providing a significant improvement over Goldberg's algorithm~\cite{goldberg1984finding} by eliminating the logarithmic factor in $n$.

(2) Charikar~\cite{charikar2000greedy} proposed to transfer the original UDS problem as a linear programming (LP) problem and developed an exact algorithm.
Given a vertex set $S \subseteq V$, let $E(S)$ be the edge set induced by $S$, i.e., $E(S) = \{u,v \in S, uv \in E\}$.
Let $x_v$ and $y_e$ be the variables assigned to the edge $e$ and vertex $v$, respectively, where $x_v = 1/|S|$ indicates that $v$ is included in $S$, and $y_e = 1/|S|$ denotes $e$ is in $E(S)$.
Then, the original UDS problem can be described as follows.
\begin{equation}
	\label{or-lp}
	\begin{aligned}
		& \max \sum_{e \in E} y_e \quad s.t.\\
		 & y_e \leq x_u, x_v, \ \forall e = uv \in E;\quad\sum_{v \in V} x_v \leq 1\\
		& x_v, y_e \geq 0, \ \forall e\in E, \forall v\in V
	\end{aligned}
\end{equation}

Based on the equation above, we can construct an optimal solution for the DSD problem by first ordering the \(y_i\) values in non-increasing order.
Let \((y_1, \ldots, y_n)\) be the variables so ordered. 
Then, we find the prefix \((y_1, \ldots, y_k)\) whose corresponding induced subgraph \(H\) achieves maximum density, for any \(k \in [2, n]\).
It can be proved that \(H\) is the DS.
The overall time complexity of the LP-based algorithm is \(O(n^4)\).
It can also be proved that the optimal solution to the LP problem is a convex combination of integral solutions~\cite{balalau2015finding}.

(3) Fang et al.~\cite{fang2019efficient} proposed an exact algorithm namely \texttt{CoreExact}, which exploits the $k$-core to improve the efficiency.
Given an undirected graph $G$, the $k$-core is the maximal subgraph in which each vertex's degree within the subgraph is at least $k$.
The core number of a vertex $v \in V$ is the largest $k$ that enables a $k$-core containing $v$, and the maximum core number among all vertices is denoted as $k_{max}$.

On the basis of \texttt{Exact}, the authors first proposed a tighter upper bound for $g$, by replacing the original $d_m$ with $k_{max}$. Since $k_{max}$ is much smaller than $d_m$, the number of binary searches is significantly reduced.
Secondly, to reduce the overhead of reconstructing the flow network, the authors proved that the optimal solution is contained in a certain $k$-core, so the DS can be located in the $k$-core through the lower bound of $g$.
Specifically, a lower bound of the maximum density is first obtained by computing the density of the remaining subgraphs during the core decomposition and further tightening it by pruning strategies. Due to the nested property of $k$-core, the vertices with smaller core numbers in the remaining graph can be continuously removed in the search process to gradually reduce the size of the flow network.

\begin{figure}[h]
	\centering
	\includegraphics[width=0.6\linewidth]{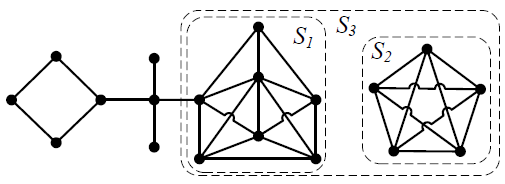}
	\caption{An example of the core-based algorithm \cite{fang2019efficient}.}
	\label{fig:coreexact}
\end{figure}

For example, for the graph in Fig. \ref{fig:coreexact}, its $k_{max}$ is 4. In the core decomposition process, the residual subgraph with the highest density is $S_3$ (its density is $25/12$), so the lower bound of $g$ is 3. Thus, the DS can be located in the 3-core.
After that, the DSs in the connected components in the 3-core, i.e., $S_1$ and $S_2$, are computed one by one by {\tt Exact} algorithm.
The worst-case time complexity of \texttt{CoreExact} is consistent with that of \texttt{Exact}. However, due to the reduced search space, \texttt{CoreExact} runs much faster than \texttt{Exact} in practice.

{\color{black}
{\bf Discussion.} Exact algorithms lay the theoretical foundation of the UDS problem by characterizing optimal solutions through flow formulations and linear programming. Flow-based methods offer strong optimality guarantees but incur high computational overhead, while LP-based approaches primarily provide structural insights and are more amenable to relaxation, making them better suited for approximation than for exact computation. In addition, core-based refinements demonstrate that exploiting intrinsic graph structure can substantially reduce the practical cost of exact algorithms.

Despite these advances, the reliance of exact methods on repeated max-flow or LP solving remains a major obstacle to scalability on large graphs. This limitation directly motivates the development of approximation algorithms, which relax flow and LP formulations to achieve better efficiency while preserving the key structural ideas, and further inspires extensions to dynamic maintenance settings.
}

\subsection{Approximation solutions}
As discussed before, the exact solutions cannot process large-scale graphs due to their prohibitive time complexities, so researchers have developed many approximation algorithms, which improve efficiency by trading the accuracy.
In this paper, the approximation ratio of the algorithm is defined as the ratio of the density of the DS to the actual density of the returned subgraph.
For instance, if the density of the returned subgraph is at least half of the maximum density, the algorithm is considered a 2-approximation algorithm, meaning it guarantees an approximation ratio of 2.

In the realm of approximation solutions:
For a 2-approximation ratio, Charikar~\cite{charikar2000greedy} introduced a greedy algorithm with a worst-case time complexity of \(O(m+n)\), while Fang et al.~\cite{fang2019efficient} proposed a \(k\)-core-based algorithm with better practical efficiency despite its time complexity of \(O(m + n)\).
{\color{black}For approximation ratios less than 2, Boob et al.~\cite{boob2020flowless} presented a greedy algorithm named \texttt{Greedy++} with a time complexity of \(O(\frac{\Delta(G) \log m}{\rho^* \epsilon^2} m\log n)\). Harb et al.~\cite{harb2022faster} proposed an LP-based algorithm with a worst-case time complexity of \(O(m \frac{\sqrt{m \Delta(G)}}{\epsilon})\). If \(\rho^* > \sqrt{\frac{\Delta(G)\log^2 n \log^2m}{m \epsilon^2}}\), \texttt{Greedy++} offers a more favorable worst-case time cost.
In contrast, when $\rho^*$ is very small, its running time may degrade and can be less competitive than exact flow-based methods, although such cases are typically uncommon in practice.
}

(1) Charikar~\cite{charikar2000greedy} first proposed a 2-approximation algorithm. The main idea of the algorithm is to continuously remove vertices with the smallest degrees to obtain subgraphs with large average degrees.
The method is based on the peeling paradigm and is recorded as \texttt{PeelApp}. Specifically, given an undirected graph with $n$ vertices, the vertex with the smallest degree in the graph is removed each time, and then the density of the remaining graph is computed. After removing $n$ vertices, the one with the largest density among all subgraphs is returned.
It is proved that the returned subgraph is a 2-approximation solution to the original UDS problem, and the time complexity of \texttt{PeelApp} is $O(m+n)$.

Tsourakakis et al.~\cite{tsourakakis2019novel} proved that the USD problem with negative edges is NP-hard.
They also showed that {\tt PeelApp}, which iteratively removes vertices with the smallest weighted degree, can approximately solve the problem. 
The resulting density is guaranteed to be at least $\frac{1}{2}(\rho^* - \Delta)$, where  $\rho^*$ is the optimal density and $\Delta$ is the upper bound on the negative degree among all vertices.

(2) Fang et al.~\cite{fang2019efficient} improved \texttt{PeelApp} by exploiting $k$-core. 
Since the $k_{max}$-core is a 2-approximation solution to the UDS problem, they designed an efficient algorithm called \texttt{CoreApp} to compute the $k_{max}$-core.
Specifically, the algorithm first arranges all vertices in the graph in non-ascending order of degrees, then selects the vertices whose degrees are greater than a certain value, and finally computes the $k$-cores of the subgraph induced by these vertices. 
If the obtained $k_{max}$ is greater than the maximum value of the vertex degree in the remaining graph, then it is the $k_{max}$ of the entire graph. Otherwise, it repeats the above steps on a larger graph by adding vertices with smaller degrees until the number of vertices is twice the original graph.
The worst-case time complexity of \texttt{CoreApp} is $O(n+m)$, but practically it runs much faster than {\tt PeelApp} because of the reduced size of the graph accessed.

To efficiently compute the DS in large-scale graphs, Luo et al.~\cite{luo2023scalable} proposed a parallel approximation algorithm to obtain $k_{max}$-core.
Specifically, the algorithm follows the h-index-based core decomposition approaches~\cite{sariyuce2018local} to iteratively compute core numbers of the vertices. 
The difference is that the authors prove the convergence condition of $k_{max}$-core.
Therefore, the algorithm computes the core numbers of vertices in the $k_{max}$-core, and avoids redundant computation for all the other $k$-cores, thereby improving the efficiency significantly.
Besides, the algorithm has good parallelism due to the locality of computation between vertices.
The time complexity of the algorithm is $O(t\cdot m)$, where $t$ is the number of iterations and $t\ll k_{max}$ in practice~\cite{sariyuce2018local}.

All the approximation algorithms above can only find solutions whose approximation ratios are at least 2.0.
To further improve the quality of returned approximation solutions, some researchers have designed algorithms with approximation ratios less than 2.
One direction is to approximate the positive LPs by numerical methods. In a positive LP, all the coefficients, variables, and constraints are non-negative, which is alternatively known as Mixed Packing and Covering LPs.

(3) Bahmani et al.~\cite{bahmani2014efficient} proposed an algorithm based on MapReduce with an approximation of $1+\epsilon$ by solving the dual LP of Eq. (\ref{eq:dual-lp}), which is described as follows.
\begin{equation}
	\label{eq:dual-lp}
	\begin{aligned}
		& \min \  D\ s.t.\\
		 \  \ & f_e(u) + f_e(v) \geq 1,\ f_e(u), f_e(v) \geq 0, \ \forall e = uv \in E\\
		& \sum_{v \in e} f_e(v) \leq D, \ \forall v \in V.
	\end{aligned}
\end{equation}
In this dual LP, each edge $e = uv$ has a load of 1, which it wants to assign to its endpoints: $f_e(u)$ and $f_e(v)$ such that the total load on each vertex is at most $D$.
The objective is to find the minimum $D$ for which such a load assignment is feasible.
Then, a ($1+\epsilon$) approximation solution to the problem is obtained by bounding the width of the problem and solving it using the multiplicative weights update framework~\cite{arora2012multiplicative,plotkin1995fast}.
{\color{black}
The width of an LP measures the maximum relative contribution of any variable to a constraint; in Eq. (4), it corresponds to the unit contribution of each $f_e(v)$ to vertex load constraints, yielding a constant width.
}
The time complexity of the algorithm is $O(\frac{m\log n}{\epsilon^2})$, where $\log n/\epsilon^2$ is the number of rounds in MapReduce.

Su and Vu~\cite{su2020distributed} adopted the same technique and proposed a distributed algorithm with an approximation ratio of ($1+\epsilon$). It adopts the acceleration method for solving positive LP~\cite{boob2019faster}, with a time cost of $\tilde{O}(m \Delta /\epsilon)$, where $\Delta$ is the maximum degree in the input graph, and $\tilde{O}$ hides the coefficient of $\log n$.

(4) Boob et al.~\cite{boob2020flowless} proposed the \texttt{Greedy++} algorithm, based on \texttt{PeelApp}.
The algorithm takes a graph $G$ and an integer $T$ as input, where $T$ is the number of iterations of the algorithm. 
The main idea of the algorithm is to obtain DS by iteratively removing the vertex with the smallest load, where the load of vertex $u$ in each iteration is the sum of its induced degree and the load of $u$ in the previous iteration. 

Specifically, it initializes the load of all vertices as 0. In each iteration, it finds the vertex with the smallest load in the current graph, updates the load of all vertices, and removes $u$ and the corresponding edges from $G$.
After $T$ iterations, it returns the subgraph with the largest density in all subgraphs as the final result. The time complexity of this algorithm is $O((m+n)\cdot min(\log n,T))$, and its approximation ratio is $1+1/\sqrt{T}$. Note that if $T=1$, the algorithm reduces to \texttt{PeelApp}.
Recently, \cite{chekuri2022densest} have proved that {\tt Greedy++} converges to a $(1+ \epsilon)$-approximation in $O(\frac{\Delta(G)\log m}{\rho^* \epsilon^2})$ iterations where $\rho^*$ is the optimum density and $\Delta(G)$ is the maximum degree of $G$. 
They also demonstrated that the algorithm converges to an approximation ratio of $(1+\epsilon)$ for any supermodular density function.
Xu et al.~\cite{xu2023efficient} experimentally evaluated the performance of \texttt{Greedy++} and \texttt{CoreApp}, demonstrating that \texttt{CoreApp} has superior efficiency, while \texttt{Greedy++} achieves a lower approximation ratio.

(5) Harb et al.~\cite{harb2022faster} proposed an algorithm based on the dual of Charikar's LP relaxation. The dual LP is as follows.
\begin{equation}
\label{dual-lp_m}
\begin{aligned}
&\min \ \max_{u \in V} \; b_u \quad s.t.\\
&\sum_{v \in \delta(u)} x_{uv} = b_u,\ \forall u \in V,\\
&\qquad x_{uv}+x_{vu}=1,\ \forall \{u,v\} \in E,\quad x_{uv}, x_{vu}, b_u \ge 0
\end{aligned}
\end{equation}

Eq.~(\ref{dual-lp_m}) can be viewed as orienting each edge fractionally towards $u$ and $v$, the orientations induce loads at the vertices, and the goal is to find an orientation that minimizes the maximum load on the vertices. 
This LP can be solved by some iterative algorithms such as MWU~\cite{su2020distributed} and Frank-Wolfe method~\cite{danisch2017large}.
This dual LP can be transformed into an unconstrained optimization problem that minimizes $f(x)+h(x)$, where $f(x)$ is a convex function and $h(x)$ has proximal mapping that is easy to compute.
This problem can be solved by the proximal gradient method. Specifically, in the $t$-th iteration, the minimal $x^{(t)}$ is guessed, then the gradient of $f$ is calculated and shifted slightly.
To make the new guess feasible, the proximal mapping is used to project the new guess to a feasible solution.
The authors then employ an accelerated proximal gradient method that incorporates Nesterov-like momentum terms~\cite{nesterov1983method}     in the projection step, resulting in faster results (both theoretically and practically). This method is also called the FISTA method~\cite{beck2009fast}.
The authors show that the method converges to an $\epsilon$-additive approximate local decomposition vector through $O(\frac{\sqrt{m \Delta(G)}}{\epsilon})$ iterations at most, with each iteration taking $O(m)$ time.

(6) Chekuri et al. \cite{chekuri2022densest} presented a flow-based approximation algorithm for the original UDS problem. Compared to flow-based exact algorithms, it does not need to compute the exact max-flow.
Specifically, the algorithm only needs to perform partial max-flow computations with certain iterations of blocking flows based on Dinic's algorithm \cite{dinitz2006dinitz}. \cite{chekuri2022densest} proved that this algorithm can give the $(1+\epsilon)$-approximation result within $\tilde{O}(\frac{m}{\epsilon})$ time cost.

Note that Bahmani et al.~\cite{bahmani2012densest} proposed an approximation algorithm for the UDS problem, achieving an approximation ratio of $2(1+\epsilon)$ where $\epsilon > 0$. 
This algorithm is tailored for MapReduce and streaming settings. Similar to \texttt{PeelApp}, it removes batches of low-degree vertices per iteration instead of peeling a single vertex.
In each pass, all vertices with degree at most $2(1+\epsilon)\rho$ are removed, where $\rho$ is the current subgraph density. The densest subgraph encountered across all iterations is returned.
The authors proved that the output achieves a $2(1+\epsilon)$-approximation to the UDS problem. 
The algorithm runs in $O(\frac{m\log n}{\epsilon})$ time, over $O(\frac{\log n}{\epsilon})$ passes of $O(m)$ each.

{\color{black}
{\bf Discussion.} 
Approximation algorithms trade exact optimality for provable guarantees, enabling scalable UDS computation on large graphs. Two dominant paradigms exist: peeling-based methods, which iteratively remove low-contribution vertices, and LP-inspired methods, which approximate fractional relaxations via iterative or numerical schemes.
Peeling-based approaches are attractive for their simplicity and near-linear complexity and underpin many dynamic and streaming algorithms. LP-based approximations, in contrast, provide a path to $(1+\epsilon)$ guarantees and motivate flowless and load-balancing formulations that extend naturally to evolving graphs.
Notably, many approximation algorithms support incremental and localized updates, making them natural building blocks for dynamic and streaming maintenance, discussed next.
}

\subsection{DS maintenance solutions}

Dynamic DSD methods differ mainly in what information is maintained under graph updates.
Peeling-based methods maintain degree-based levels that approximate the static peeling process, allowing the solution to be updated locally but typically requiring the current graph to be retained ~\cite{epasto2015efficient}.
Sampling-based methods instead maintain a sparsified representation or sketch of the graph, substantially reducing space and update cost at the expense of additional query or post-processing~\cite{bhattacharya2015space,esfandiari2015applications,mcgregor2015densest}.
LP-dual-based methods view DSD as an edge-load/orientation problem and restore local load invariants after each update, achieving near-optimal approximation with polylogarithmic update time~\cite{sawlani2020near,chekuri20221}.

Specifically, the algorithm \texttt{PeelApp} proposed by Bahmani et al.~\cite{bahmani2012densest} is also suitable for streaming graphs; Das Sarma et al.~\cite{das2012dense} adopted the same techniques to maintain the DS on the distributed congest model.
Similarly, the ($1+\epsilon$)-approximation algorithms on static graphs proposed by Bahmani et al.~\cite{bahmani2014efficient} can be applied to dynamic graphs.


Subsequently, Bhattacharya et al.~\cite{bhattacharya2015space} developed a 1-pass streaming algorithm with a ($2+\epsilon$) approximation ratio by designing a subtle data structure.
They also provided a fully dynamic DS algorithm with a ($4+\epsilon$) approximation ratio, featuring an amortized update time and space complexities of $O(poly(\log n,\epsilon^{-1}))$ and $\tilde{O}(n)$, respectively.
Their approach effectively addresses the challenge of applying $l_0$ samplers~\cite{cormode2014unifying}.
Epasto et al.~\cite{epasto2015efficient} proposed a fully dynamic DSD algorithm with an approximation of ($2+\epsilon$) based on \texttt{PeelApp}. The amortized time of each update operation of the algorithm is $O(\log^2 n/\epsilon^2)$, where edges are randomly deleted.
Esfandiari et al.~\cite{esfandiari2015applications} improved the algorithms presented in~\cite{bhattacharya2015space} by introducing a $(1+\epsilon)$-approximation algorithm using $\tilde{O}(n)$ space, which employs min-wise independent hashing combined with fast multi-point polynomial evaluation.
McGregor et al.~\cite{mcgregor2015densest} proposed a single-pass algorithm with an approximation ratio of $(1+\epsilon)$. It uses $O(\epsilon^{-2} n \text{polylog}(n))$ space and processes each stream update in $polylog(n)$ time, and incurs $poly(n)$ post-processing time.
Saurabh and Wang~\cite{sawlani2020near} proposed a fully dynamic DSD algorithm with an approximation ratio of $1+\epsilon$. The algorithm is also implemented by solving the dual problem of LP, which transforms the problem into a problem of assigning edge loads to associated vertices to minimize the maximum load between vertices. In the worst case, the running time of each update operation is $O(poly(\log n,\epsilon^{-1}))$.

\setlength{\dbltextfloatsep}{3pt plus 1pt minus 1pt}







\begin{table*}[t]
\centering
\caption{
Comparison of representative dynamic/streaming DSD algorithms. ``Solution Access'' indicates whether the density value or the explicit subgraph is maintained online, or requires additional recovery.
}
\label{tab:dynamic_tradeoff}

\resizebox{\textwidth}{!}{
\begin{tabular}{l|l l l l l l}
\hline
Method
& Update Model
& Principle
& Approx.
& Update Time
& Solution Access
& Space \\
\hline
\hline

Bhattacharya et al.~\cite{bhattacharya2015space}
& Dynamic stream
& Decomp. + sampling
& $4+\epsilon$
& $\tilde{O}(1)$ amort.
& Density: $\tilde{O}(1)$
& $\tilde{O}(n)$ \\

Epasto et al.~\cite{epasto2015efficient}
& Adv. ins.; random del.
& Peeling/levels
& $2(1+\epsilon)$
& polylog amort.
& Subgraph maintained
& $O(n+m)$ \\

McGregor et al.~\cite{mcgregor2015densest}
& Dynamic stream
& Sampling/sketching
& $1+\epsilon$
& polylog worst-case
& $\mathrm{poly}(n)$ recovery
& $O(\epsilon^{-2}n\,\mathrm{polylog}n)$\\

Sawlani--Wang~\cite{sawlani2020near}
& Fully dynamic
& LP dual/orientation
& $1+\epsilon$
& polylog worst-case
& Density: $O(1)$
& -- \\

Chekuri--Quanrud~\cite{chekuri20221}
& Fully dynamic
& Dynamic orientation
& $1+\epsilon$
& polylog (amort./worst-case)
& Density: $O(1)$
& linear \\

\hline
\end{tabular}
}
\end{table*}

{\bf Discussion.}
The applicability of these methods depends strongly on the update model and the required time--space trade-off. Most existing guarantees are derived for single-edge updates: Epasto et al.~\cite{epasto2015efficient} allow adversarial insertions but assume random deletions, whereas the dynamic-streaming and orientation-based methods support both insertions and deletions. Explicit batch-update guarantees are generally unavailable, and we therefore do not infer batch performance from single-update bounds. The reported update bounds also have different semantics. Amortized bounds control the average cost over an update sequence, whereas worst-case bounds guarantee the latency of every update. Sampling-based methods typically achieve $\tilde O(n)$ space and fast stream updates but may require expensive post-processing, while peeling-based methods may retain $O(n+m)$ graph state. Recent orientation-based methods provide near-optimal approximation with worst-case polylogarithmic updates, and Chekuri and Quanrud~\cite{chekuri20221} further reduce the space to linear. Table~\ref{tab:dynamic_tradeoff} reports only guarantees explicitly established in the original works.

\subsection{Variants of the original UDS problem}
%
%
{\color{black}

Beyond the classical formulation, numerous UDS variants on undirected graphs have been proposed to capture richer density notions or incorporate application-driven constraints (see Table~\ref{tab:models_undirected}).
In this survey, \emph{densest subgraph} refers to the classical maximum-density problem, while terms such as \emph{top-$k$ densest subgraphs} and \emph{locally densest subgraphs} denote extended variants with additional objectives or constraints.
Existing variants can be broadly grouped into two categories: (i) redefining density, e.g., via clique-, pattern-, or motif-based measures; and (ii) imposing structural or combinatorial constraints, such as size, diversity, or locality.

Despite this diversity, most approaches build on the same algorithmic foundations as UDS, including flow-based methods, LP relaxations, and peeling or core-based strategies, adapted to new density definitions or constraints.

}

\subsubsection{Clique-density-based DSD}
We first introduce the definition of clique-density. 

\begin{definition}[\textbf{$k$-clique}~\cite{danisch2018listing,li2020ordering}]
\label{def:kclique}
A $k$-clique is a complete graph with $k$ vertices, where there is an edge between every pair of vertices.
\end{definition}

\begin{definition}[\textbf{Clique-density}~\cite{tsourakakis2015k,samusevich2016local,fang2019efficient}]
\label{def:clique-density}
Given an undirected graph $G$=$(V,E)$ and a $k$-clique $\Psi$ with $k \geq 2$, its clique-density $w.r.t.$ $\Psi$ is defined as
 $\rho(G, \Psi) = \frac{{u(G, \Psi)}}{{|V|}}$,
where $u(G, \Psi)$ is the number of clique instances of $\Psi$ in $G$.
\end{definition}

Clearly, the clique-density is an extension of edge-density, since when $\Psi$ is an edge, it reduces to the edge-density.
Since clique-density-based densest subgraph (CDS) has the same properties as edge-density-based densest subgraph (EDS), the algorithms of the original UDS problem could be extended to solve the CDS problem.

Mitzenmacher et al.~\cite{mitzenmacher2015scalable} proposed an exact algorithm for CDS discovery, which is also based on a flow network to search the CDS. 
Then the CDS can be obtained by the flow-based exact algorithm. The time complexity of the algorithm is $O(m\alpha(G)^{k-2}+(n+c_k)^2)$ where $\alpha(G)$ is the arboricity of $G$ and $c_k$ is the number of $k$-clique instances in $G$.
To further improve the efficiency, the authors proposed a sampling-based approximation algorithm. This method sparsifies the graph through sampling and identifies the CDS on the sparse graph using an exact algorithm. By setting the appropriate sampling probability, the algorithm can obtain a $(1+\epsilon)$-approximate solution to the CDS problem where $\epsilon>0$.



Tsourakakis~\cite{tsourakakis2015k} studied the triangle-density-based DS problem in undirected graphs. He proposed a new exact algorithm based on supermodularity besides flow networks. 
Specifically, let $t(S)$ be a function that returns all triangles in the induced subgraph of the specified vertex set $S$. The author proved that $f(S) = t(S) -\alpha|S|$ is supermodular, where $\alpha$ is the guessed density of the CDS. According to supermodularity, the algorithm initializes the upper and lower bounds of $\alpha$ and computes the triangle-density-based DS by binary search. In each iteration, the algorithm takes $G$ and $\alpha$ as input and maximizes $f_\alpha$ using Orlin-Supermodular-Opt~\cite{orlin2009faster}. The time complexity of the exact algorithm is $O(\log n(n^5m^{1.4081}+n^6))$.
Subsequently, the author proposed a peeling-based 3-approximation algorithm. 
Specifically, it iteratively deletes the vertex of the minimum number of triangles in the graph and returns the subgraph with the largest triangle density.

Fang et al.~\cite{fang2019efficient} proposed the concept of $(k, \Psi)$-core based on $k$-core. Given an integer $k$ and an $h$-clique $\Psi$, all vertices in $(k, \Psi)$-core are contained by at least $k$ instances of $\Psi$. Based on $(k, \Psi)$-core, \texttt{CoreExact} and \texttt{CoreApp} can provide exact and $|V_\Psi|$-approximation solutions for the CDS problem, where $|V_\Psi|$ is the number of vertices in $\Psi$.

Sun et al.~\cite{sun2020kclist++} proposed a more efficient CDS algorithm. 
The main idea of the algorithm is to introduce $k$ variables for each $k$-clique and iteratively update them to find the CDS. The algorithm is an instance of the Frank-Wolfe algorithm~\cite{jaggi2013revisiting}.
Specifically, for each $k$-clique $C$ in $G$, assign a variable $\alpha_u^C$ to each vertex $u$ in $C$, initialize it to $\frac{1}{k}$, and assign a variable $r(u)$ to each vertex $u$ in $G$, which is initialized to the sum of all $\alpha_u^C$ such that $C$ contains $u$. 
For each $k$-clique $C$, let $x$ be the minimum value of $r(u)$ in all vertices of $C$. Then define a variable $\hat{\alpha}^C_u$ for each vertex $u$, if $\alpha_u^C = x$, then $\hat{\alpha}^C_u = 1$, otherwise 0. In each iteration of the algorithm, the value of $\alpha_u^C$ is updated by a convex combination of $\alpha_u^C$ in the previous iteration and $\hat{\alpha}^C_u$. 
The induced subgraph of the corresponding vertex set $S$ with the largest values of $r$ is the CDS of $G$.
Besides, the authors proposed a non-gradient descent method named KCLIST++ to compute the CDS and reduce memory consumption based on the $k$-clique enumeration algorithm KCLIST~\cite{danisch2018listing}. 
Specifically, the algorithm initializes $r(u)$ of each vertex in $G$ to 0 and sequentially processes all $k$-clique, that is, adding 1 to $r(u)$ of the vertex with the smallest $r$ among all vertices of the $k$-clique each time, after $T$ rounds of iterations, each $r(u)$ is divide by $T$.
The time cost of the algorithm is $O(T \cdot k\cdot m \cdot (\frac{c}{2})^{k-2})$, where $c$ is the maximum core value of the graph.

KCLIST++ struggles to scale with large graphs due to its repeated enumeration of all $k$-cliques in each iteration.
To address this issue, He et al.~\cite{he2023scaling} proposed SCTL, which accelerates the $k$-clique enumeration process by building an index structure. This approach leverages the succinct clique tree from PIVOTER~\cite{jain2020power}, the state-of-the-art algorithm for $k$-clique counting. 
The time complexity of SCTL is $O(n\cdot 3^{c/3} + T k \Psi_k(G))$, where $\Psi_k(G)$ denotes the number of $k$-cliques in the graph.
However, SCTL still faces the challenge of enumerating all $k$-cliques in the worst case for each iteration, as it updates vertex weights based on these cliques. 
This results in the running time being proportional to the number of $k$-cliques in the graph.
To overcome this limitation, Zhou et al.~\cite{zhou2024counting} introduced the KCCA algorithm, which updates vertex weights more efficiently by counting only the $k$-cliques that contain the vertex with the minimum weight.
The time complexity of KCCA is $O(T\cdot n\cdot 3^{c/3} \cdot c\log c)$. KCCA significantly improves performance compared to clique enumeration-based algorithms.

In addition to $k$-clique, edge-density can also be extended to pattern-density by replacing the number of edges with the number of instances of a given pattern. 
A pattern is a small graph composed of small parts of vertices, also known as a motif or higher-order structure.
Fang et al.~\cite{fang2019efficient} studied the pattern-density-based DSD problem. Similar to CDS, this problem can also be solved by \texttt{CoreExact} and \texttt{CoreApp}, where the approximation of \texttt{CoreApp} is $|V_P|$ for a given pattern $P$, and $|V_P|$ is the number of vertices in $P$.

{\color{black}
{\bf Discussion.} Clique- and pattern-density variants generalize edge density by counting higher-order structures.
Although exact solutions often rely on flow or supermodular optimization and suffer from high complexity, practical algorithms increasingly adopt peeling, core-based, or LP-relaxation techniques.
These methods preserve the structural intuition of UDS while extending its applicability to higher-order graph patterns.
}

\subsubsection{Densest $k$-subgraph}

Many studies attempt to impose constraints on the result of DSD. A typical one is the dense $k$-subgraph (DKS) problem, which aims to find the DS with $k$ vertices in the graph. 
It can be regarded as a generalization of the maximum clique problem, which belongs to a class of well-known problems called fixed cardinality problems.

Bourgeois et al.~\cite{bourgeois2013exact} proposed an exact algorithm for the DKS problem. It divides the vertex set $V$ of $G$ into two subsets $V_1$ and $V_2$. 
For each $j \in [0, k]$, it enumerates the subset $A_1$ of size $j$ in $V_1$, and finds the subset of size $k-j$ in $V_2$ such that the induced subgraph composed of $A_1$ and $A_2$ has the largest number of edges.
The algorithm has exponential time complexity since enumerating all subsets of $V_1$ takes $O(2^{V_1})$ time. 
Besides, several efficient approximate solutions have been developed.
Billionnet et al.~\cite{billionnet2008deterministic} proposed an algorithm with an approximation ratio of $\frac{9n}{8k}$; Feige et al.~\cite{feige2001dense} proposed an algorithm with an approximation ratio of $O(n^{\frac{1}{3}})$; Bhaskara et al.~\cite{bhaskara2010detecting} proposed an $O(n^{\frac{1}{4}-\epsilon})$-approximation solution, where $\epsilon>0$.
Bourgeois et al.~\cite{bourgeois2013exact} proposed a series of approximation algorithms, where the approximation ratio depends on the time complexity of the algorithm.

Asahiro et al.~\cite{asahiro2000greedily} studied the DKS problem on weighted graphs. Since it is an NP-complete problem, the authors proposed a greedy-based algorithm that iteratively removes the vertices with the smallest weighted degree in the graph until there are $k$ vertices remaining. 
The weighted degree of a vertex is the sum of the weights of all edges connected to the vertex, which reduces to its degree in unweighted graphs. The approximation of the greedy algorithm is related to the number of vertices in the graph, i.e., if $\frac{n}{3} \leq k \leq n$, the range of the approximation $r$ is $[(1/2+n/2k)^2-O(n^{-1/3}),(1/2+n/2k)^2+O(1/n)]$, and if $k < \frac{n}{3}$, $r \in [2(n/k-1)-O(1/k),2(n/k-1)+O(n/k^2)]$.

Anderden and Chellapilla~\cite{andersen2009finding} studied the problem of finding dense subgraphs with upper and lower bound constraints on the graph size; that is, finding the DS with at least $k$ vertices (DALKS) and the DS with at most $k$ vertices (DAMKS) in the graph, both of which are variants of the DKS problem.
For DALKS, the authors proposed a peeling-based method similar to \texttt{PeelApp}, which continuously removes the vertices with the lowest degree in the graph and calculates the density of the remaining subgraphs, and finally returns the DS that has at least $k$ vertices.
The approximation ratio of the method is 3 and the time cost is $O(m+n)$.
For DAMKS, the authors proved that it is NP-complete and hard to approximate, and also showed that if there is a DAMKS algorithm with an approximation of $r$, then there must be a polynomial algorithm for the DKS problem with an approximation ratio of $8r^2$.

\subsubsection{Top-$k$ DSD}
Galbrun et al.~\cite{galbrun2016top} proposed the top-$k$ overlapping DSD problem, which aims to find $k$ subgraphs with high total density in the graph, and overlaps between subgraphs are allowed. 
The authors transform the problem into a max-sum diversification problem: given an integer $k$ and a set $U$, find a subset $S$ of size $k$ in $U$ that maximizes $f(S)+\lambda \sum_{x,y \in S}d(x,y)$, where $f$ is a monotonic function of a subset of $U$, $d$ is the distance function between two elements in $U$, and $\lambda$ is a parameter. This problem can be solved by the greedy algorithm framework proposed by Borodin et al.~\cite{borodin2012max}, which only needs to regard $U$ as the power set of $V$. Specifically, for the power set $U$, initialize $S$ to be empty, and iteratively add the subgraphs that do not belong to $S$, so that the current marginal gain is maximized until there are $k$ subgraphs in $S$. The margin gain added to each subgraph is calculated by Charikar's algorithm~\cite{charikar2000greedy}. The approximation of the algorithm is 10. The time complexity is $O(k(m+n(t+k)))$, where $t = min\{2^k,n\}$. 
Dondi et al.~\cite{dondi2021novel} studied the top-$k$ connected DS problem in dual networks and proposed a heuristic algorithm. The same technique can also be used to obtain a 2-approximation algorithm when the value of $k$ is less than the number of vertices~\cite{dondi2021top}.

Nasir et al.~\cite{nasir2017fully} studied the problem of top-$k$ DS maintenance. Since dense subgraphs are usually composed of vertices with relatively large degrees, they reduced the search space by considering vertices with high degrees and dividing the graph into multiple subgraphs, which can be maintained by local updates.
To achieve this, the authors first proposed a data structure called snowball, in which all vertices are connected and the core number of the vertex is equal to the largest $k$-core of the snowball. The supergraph containing multiple snowballs and the edges connecting them is then stored in the data structure, namely bag. Then the core maintenance method is applied to dynamically maintain the snowball in bag, where the $k$ subgraphs are the top-$k$ DS. Since the DS is $k_{max}$-core, which is a 2-approximation solution to DSD, the approximation of the entire algorithm is $2k$.

{\color{black}
{\bf Discussion.} 
Size-constrained and cardinality-bounded variants, such as DKS, DALKS, and DAMKS, highlight the trade-off between density maximization and combinatorial constraints.
While many of these problems are NP-hard and difficult to approximate, greedy peeling and core-based strategies remain effective heuristics, reinforcing their central role in constrained DSD variants.
}

\subsubsection{Maximum total density DSD}
Balalau et al.~\cite{balalau2015finding} studied the maximum total density DSD, which aims to find out the maximum total density of at most $k$ subgraphs while satisfying that the Jaccard coefficient between the vertex sets of any two subgraphs is not greater than a given threshold $\alpha$. 
The authors prove that the problem is NP-hard. To solve this problem, the authors first define minimal DS; that is, the DS with the least number of vertices. 
Then an algorithm is proposed to compute a minimal DS containing vertex $u$, which is based on the LP of the DS problem (Eq. (\ref{or-lp})) while adding the constraint $\sum y_e = \rho_{max}$ and maximizing the objective function $x_u$. It can be solved by using the exact algorithm or the approximate algorithm proposed by Charikar~\cite{charikar2000greedy}.
To find subgraphs with the maximum total density, the authors first find a minimal DS $G(V_i, E_i)$, for each vertex $v \in V_i$, count the number of vertices that are not in $V_i$ among all the neighbors of $v$ in $G$, and remove $(1-\alpha)|V_i|$ vertices with the smallest value, then repeat the above steps until $k$ subgraphs are obtained or $G$ is empty.

\subsubsection{Density-friendly graph decomposition}

Tatti and Gionis~\cite{tatti2015density} proposed density-friendly graph decomposition, a variant of $k$-core decomposition that orders subgraphs by density. 
They introduce the notion of outer density: for vertex sets $X$ and $Y$, let $E_\Delta(X,Y)$ denote edges with at least one endpoint in $X$. The outer density is defined as $d(X,Y)=|E_\Delta(X,Y)|/|X|$. 
A vertex set $W$ is locally dense if there exist no $X \subseteq W$ and $Y \cap W = \emptyset$ such that $d(X, W\!\setminus\!X) \le d(Y, W)$. The goal is to find a nested chain of locally dense subgraphs.
Let $\{B_i\}$ denote this chain, where $B_0=\varnothing$ and $B_k=V$. For each $i$, $B_i$ is the densest subgraph strictly containing $B_{i-1}$, and $d(B_i,B_{i-1}) > d(B_{i+1},B_i)$.

An exact algorithm is proposed based on \texttt{Exact}~\cite{goldberg1984finding}. Unlike \texttt{Exact}, which computes only $\alpha_1=\rho^*$, this method computes a sequence $\{\alpha_i\}$ (with $k \le n$), each corresponding to a locally dense subgraph. It initializes boundary values and recursively identifies intermediate $\alpha_i$ by checking whether adjacent subgraphs are consecutive; otherwise, new subgraphs are inserted. The time complexity is $O(n \cdot t_{\text{Flow}})$, as max-flow is invoked at most $2k-3$ times.
To improve efficiency, a 2-approximation algorithm based on \texttt{PeelApp} is proposed. It first computes a 2-approximate densest subgraph, then iteratively constructs the chain by selecting subgraphs maximizing $d(B_j,B_{j-1})$. This algorithm runs in $O(m)$ time.

Due to the high cost of exact methods, Danisch et al.~\cite{danisch2017large} proposed a more scalable approach based on the Frank-Wolfe algorithm. 
They maintain edge weights $w_e$ distributed to endpoints as $\alpha_u^e$, and define $r(u)=\sum_e \alpha_u^e$. At each iteration, $w_e$ is assigned to the endpoint with minimum $r(u)$, yielding converged $\alpha$ and $r$ after $T$ iterations.
A heuristic decomposition is then derived by sorting vertices in decreasing $r(u)$, generating candidate subsets via the PAVA algorithm~\cite{calders2014mining}, and verifying them against the exact solution. 
They also provide an error bound estimation and an $\epsilon$-approximation scheme: the decomposition is iteratively refined until the estimated error falls below a given threshold.

\subsubsection{Locally DSD}
To define the locally densest subgraph (LDS), Qin et al.~\cite{qin2015locally} first introduced a new concept, namely $\rho$-compact. Given an undirected connected graph $G$ and a nonnegative real number $\rho$, $G$ is $\rho$-compact if removing any subset $S$ of $V$ removes at least $\rho|S|$ edges from $G$.
A subgraph $g$ of $G$ is an LDS if $g$ is a maximal $\rho_g$-compact subgraph, where $\rho_g$ is the density of $g$.
Given a graph $G$ and an integer $k$, locally DSD aims to find the top-$k$ LDSes in $G$ with the largest density.
Note that a DS is also an LDS.
The authors proposed a greedy algorithm, which computes the DS of $G$ and removes a connected component in DS from $G$. If the connected component is an LDS, it is added to the result. These steps are repeated until $k$ subgraphs are obtained.
In the worst case, it needs to compute the DS and verify LDS $O(n)$ times, so the time complexity is $O(t_{\text{Flow}} \cdot (m+n)\log^2n)$.

The algorithm above needs to run the max-flow algorithm on the graph several times to find an LDS candidate and verify it, so it may not scale well on large graphs.
To alleviate this issue, Ma et al.~\cite{ma2022finding} proposed a top-$k$ LDS search algorithm based on convex programming.
Specifically, they defined a compact number for a vertex in the graph, which represents the most compact subgraph containing the vertex, and the compact number of each vertex can be obtained by solving the convex program depicted in \cite{danisch2017large}. Since the vertices in an LDS share the same compact number, the upper and lower bounds of the vertex number can be obtained through the Frank-Wolfe algorithm, and most of the vertices are filtered by the proposed pruning strategies to obtain subgraphs smaller than $k$-core. The final result is obtained by computing the min-cut of the subgraphs.
The time complexity of the algorithm is $O( (N_{FW}+N_{SG})\cdot (m + n) + N_{\text{Flow}} \cdot t_{\text{Flow}})$, where $N_{FW}$ is the number of iterations of the Frank-Wolfe algorithm, $N_{SG} \leq n$ is the number of candidates of LDSes, $N_{\text{Flow}}$ is the number of times the verified approach is called, and $t_{\text{Flow}}$ is the time complexity of min-cut computation.

\subsubsection{DS deconstruction}
There may be multiple subgraphs with the largest density in a graph, and DSD always returns one of them.
Chang and Qiao~\cite{chang2020deconstruct} studied how to efficiently output all DS or minimal DS in an undirected graph.
To organize all the DSs, the authors define and study the flow network $H$ corresponding to $\rho_{max}$. Let $f^*$ be a max flow of $H$ and $H_f^*$ be the residual graph of $H$ under $f^*$. To enumerate all DSes, the authors treat $H_f^*$ as a directed graph and decompose it into strongly connected components (SCCs). An SCC is called non-trivial if it does not contain source node $S$ and target node $T$. The authors organize all non-trivial-SCCs into an index called ds-Index. Specifically, each SCC is regarded as a super node, and the directed edge between the nodes indicates that there is a directed path between the two SCCs. Two sets are independent if they are not successors to each other. Enumerating all the independent sets in the index and the induced subgraphs of their successors can get all the DSs. An SCC without outgoing edges in ds-Index is called a black hole component, which corresponds to a minimal DS. The space complexity of ds-Index is $O(L)$, where $L$ is the sum of the sizes of all maximal DSs and $L\leq m+n$. The time complexity to query all DSs or minimal DS is $O(L)$.

{\color{black}
\subsubsection{Anchored densest subgraph search}

Dai et al.~\cite{dai2022anchored} introduced the anchored densest subgraph (ADS) problem to identify dense subgraphs that are relevant to a reference set while containing a prescribed anchor set. Given an undirected graph $G=(V,E)$, an anchor set $A\subseteq V$, and a reference set $R\subseteq V$ with $A\subseteq R$, the $R$-subgraph density of $S\supseteq A$ is defined as $\rho_R(S)=\bigl(2|E(S)|-\sum_{v\in S\setminus R}d_G(v)\bigr)/|S|$. Compared with conventional density, it penalizes high-degree vertices outside $R$, thereby favoring subgraphs closely related to the reference set.

The key algorithmic idea is a parametric minimum-cut reduction. For a candidate density $\alpha$, the condition $\rho_R(S)>\alpha$ can be transformed into a cut-capacity test. In the augmented network $G_\alpha$, the source $s$ connects to each anchor in $A$ with infinite capacity, to each vertex in $R\setminus A$ with capacity $d_G(v)$, each original edge has unit capacity, and every vertex connects to the sink $t$ with capacity $\alpha$. Accordingly, infinite capacities force anchors to remain in the solution, degree-weighted source edges penalize excluding reference vertices, original edges measure the boundary of $S$, and sink edges encode the size penalty. The optimal density can thus be found by binary search over $\alpha$ with repeated minimum-cut computations. To improve scalability, Dai et al. further developed a local flow algorithm that progressively expands the explored region around $R$, making it particularly suitable when the query region is small relative to the whole graph.

Ye et al.~\cite{ye2024efficient} observed that the global-degree penalty in $\rho_R(S)$ may overly suppress high-degree vertices even when they are well connected within the returned subgraph. They proposed the NR-subgraph density $\rho_R^+(S)=\bigl(2|E(S)|-\sum_{u\in S\setminus R}|N(u,S)|\bigr)/|S|$, which replaces global degree with internal degree and therefore better preserves internal connectivity. More importantly, the resulting problem admits an LP formulation whose dual can be expressed as a convex program, enabling efficient Frank--Wolfe optimization. Thus, the original ADS favors stronger regularization toward $R$ through an exact flow formulation, whereas NR-subgraph density favors internally cohesive solutions and more scalable convex optimization.

\subsubsection{Differential privacy DSD}

Differential privacy (DP) protects sensitive graph relationships by ensuring that the output changes only slightly when a single edge is added or removed. For DSD, the main difficulty is that the algorithm releases an entire vertex set rather than only a density value, so privacy must be incorporated into the search procedure itself.

Nguyen et al.~\cite{nguyen2021differentially} proposed sequential and parallel algorithms under edge DP. Their sequential algorithm privatizes Charikar's peeling process using the exponential mechanism. Instead of deterministically removing the minimum-degree vertex, a vertex $v$ is selected with probability proportional to $\exp(-\varepsilon' d_{S_t}(v))$, where $\varepsilon'=\varepsilon/(4\ln(e/\delta))$. The remaining privacy budget is used to privately select one candidate subgraph from the peeling sequence according to its density. By composition, the overall algorithm satisfies $(\varepsilon,\delta)$-edge DP and achieves $\rho(S)\geq \rho(S^*)/2-O(\log n/\varepsilon)$ up to logarithmic factors in $\delta$. Hence, a smaller $\varepsilon$ provides stronger privacy but incurs a larger additive density loss. Their parallel variants remove multiple low-degree vertices simultaneously, reducing synchronization rounds at the cost of larger utility loss and, for some variants, weaker worst-case efficiency.

Dhulipala et al.~\cite{dhulipala2022differential} adopted a different approach based on multiplicative-weight updates for the dual formulation of DSD. Instead of randomizing the peeling order, they inject symmetric geometric noise into local edge-load statistics and density tests, while distributing the privacy budget across multiple density guesses, update phases, and repeated tests. This yields a solution with density approximately $D^*/(1+\eta)-O(\operatorname{poly}(\log n)/\varepsilon)$, improving the multiplicative approximation from roughly $1/2$ to nearly $1$. They also studied the local DP setting, which eliminates the trusted curator but generally incurs weaker utility and higher communication cost. These methods expose three important trade-offs: privacy versus additive error, approximation quality versus computational complexity, and central versus local trust models.

\subsubsection{Densest diverse subgraphs search}

Diversity in DSD can refer to either \emph{structural diversity} among multiple returned subgraphs or \emph{attribute diversity} within a single subgraph. Structural diversity aims to avoid returning highly overlapping dense regions, whereas attribute diversity requires different vertex groups or labels to be sufficiently represented.

For structural diversity, Balalau et al.~\cite{balalau2015finding} imposed an upper bound on the pairwise Jaccard overlap among multiple dense subgraphs, explicitly controlling redundancy but making the optimization NP-hard. Galbrun et al.~\cite{galbrun2016top} instead combined subgraph density with pairwise distance in the objective, providing a softer trade-off between density and novelty. Thus, hard overlap constraints are preferable when explicit redundancy control is required, whereas distance-based formulations allow meaningful overlaps such as shared hub vertices.

For attribute diversity, Anagnostopoulos et al.~\cite{anagnostopoulos2020spectral} studied fair DSD with two vertex colors and required balanced representation in the selected subgraph. They proposed a greedy $2$-approximation under suitable conditions and a spectral approach that incorporates fairness through an orthogonality constraint. The spectral formulation is computationally attractive but its guarantees depend on structural assumptions that may be restrictive for highly skewed graphs.

Miyauchi et al.~\cite{miyauchi2023densest} generalized attribute diversity to multiple colors. Their densest diverse subgraph problem (DDSP) bounds the maximum fraction occupied by any color, i.e., $c_{\max}(S)/|S|\leq\alpha$, thereby controlling relative representation, and admits an $\Omega(1/\sqrt n)$-approximation. They also introduced the densest at-least-$\vec{k}$-subgraph problem, which requires $|S\cap V_c|\geq k_c$ for each required color and admits a $1/3$-approximation. DDSP is more suitable when diversity is expressed as a proportion, whereas the latter is preferable when explicit minimum representation of specific groups is required.

Overall, structural-diversity methods mainly diversify a \emph{set of outputs} and require only graph topology, while attribute-diversity methods enforce representation constraints \emph{within each output} and require vertex attributes. The former is more suitable for diversified top-$k$ exploration, whereas the latter is preferable for fairness-aware community discovery and applications with explicit representation requirements.

}

{\color{black}
{\bf Discussion.} Across a wide spectrum of UDS variants, ranging from alternative density definitions to constrained and application-driven formulations, a unifying pattern emerges.
Most variants extend the original UDS problem by modifying the objective or feasible set, while continuing to rely on a small set of core algorithmic ideas, including flow and LP formulations, peeling, and core-based pruning.
This consistency underscores the foundational role of classical UDS techniques and explains why advances in approximation and maintenance algorithms readily transfer to many of these variants.
}

    \section{DSD on directed graphs}
\label{sec:Directed}


{\color{black}
This section reviews the solutions for the densest subgraph problem on directed graphs.
Compared with the undirected case, directed graphs introduce two coupled vertex subsets and asymmetric degree constraints, which significantly complicate both problem formulation and algorithm design.
Similarly, we organize the discussion into exact algorithms, approximation algorithms, and dynamic maintenance algorithms.
}

\subsection{Exact solutions}
Similar to the undirected case, solutions to the DDS problem also follow two streams: 
\begin{enumerate}
    \item linear/convex programming-based solutions \cite{charikar2000greedy,ma2022convex};
    \item flow network-based solutions \cite{khuller2009finding,ma2020efficient}.
\end{enumerate}


Since DDS involves two vertex subsets \(S\) and \(T\), the ratio \(c = \frac{|S|}{|T|}\) has \(O(n^2)\) possible values, forming the main source of inefficiency in early exact algorithms.
A straightforward approach enumerates all \(c\) and applies UDS solutions, leading to time complexities of \(O(n^6)\) for LP-based methods~\cite{khuller2009finding} and \(O(n^2 \log n \cdot t_{\text{Flow}})\) for flow-based methods~\cite{charikar2000greedy}.
To address this, later algorithms reduce the number of evaluated \(c\) values.
In particular, Ma et al.~\cite{ma2020efficient,ma2022convex} propose flow- and LP-based methods with complexities \(O(k \log n \cdot t_{\text{Flow}})\) and \(O(h \cdot t_{\text{Flow}})\), respectively, where \(k,h \ll n^2\) in practice.


\subsubsection{LP-based algorithms}

Charikar \cite{charikar2000greedy} proposed the first exact DDS solution, which is based on linear programming (LP). 
For each $c=\frac{|S|}{|T|}$, the corresponding LP is formulated by Eq. (\ref{equ:charikar-lp}).
\begin{equation}
\label{equ:charikar-lp}
\begin{aligned}
\mathsf{LP}(c)\quad 
\max \quad & x_{\mathrm{sum}}=\sum_{(u,v)\in E} x_{u,v} \\
\text{s.t.}\quad 
& 0 \le x_{u,v} \le s_u,\ \ x_{u,v} \le t_v, \ \forall (u,v)\in E,\\
& \sum_{u \in V} s_u = \sqrt{c}, \quad \sum_{v \in V} t_v = \frac{1}{\sqrt{c}}.
\end{aligned}
\end{equation}
The variables in Eq. (\ref{equ:charikar-lp}) can be used to infer the DDS when $c=\frac{|S^{*}|}{|T^{*}|}$. Specifically, $s_{u}$, $t_{v}$, and $x_{u,v}$ indicate the inclusion of a vertex $u$/vertex $v$/edge $(u,v)$ in the DS according to whether the variable value is larger than 0 when $c=\frac{|S^{*}|}{|T^{*}|}$. To find the DDS, Charikar's algorithm needs to solve $O(n^{2})$ LPs with LP solvers.

To reduce the number of LPs to be solved, Ma et al. \cite{ma2022convex} introduced a relaxation, $a+b=2$, to the LP formulation of the DDS problem, as shown in Eq. (\ref{equ:ma-lp}). 
Comparing Eq. (\ref{equ:ma-lp}) with Eq. (\ref{equ:charikar-lp}), we can find that the two formulations are identical if we restrict $a=1$ and $b=1$. By introducing the relaxation, Ma et al. \cite{ma2022convex} managed to build the connection between each LP and the DDS. Based on the connection, they developed a divide-and-conquer strategy to reduce the number of LPs to be solved. 


\begin{equation}
\label{equ:ma-lp}
\begin{aligned}
\mathsf{LP}(c)\quad 
\max \quad & x_{\mathrm{sum}}=\sum_{(u,v)\in E} x_{u,v} \\
\text{s.t.}\quad 
& 0 \le x_{u,v} \le s_u,\ \ x_{u,v} \le t_v,\ \forall (u,v)\in E,\\
& \sum_{u \in V} s_u = a\sqrt{c}, \quad 
  \sum_{v \in V} t_v = \frac{b}{\sqrt{c}}, \quad 
  a + b = 2.
\end{aligned}
\end{equation}

To efficiently extract the DDS candidate from each specific LP, Ma et al. \cite{ma2022convex} derived the dual program of the LP and designed a Frank-Wolfe algorithm variant to optimize the dual program. They also designed early stop strategies, where max-flow computation is performed on a small subgraph, to extract the DDS candidate from the feasible solution of the dual program instead of the optimal solution. 

{\color{black}
LP-based formulations play a dual role in DDS research.
Although solving them exactly is computationally expensive, they reveal important structural properties of optimal solutions and provide a natural foundation for convex relaxations.
These insights are later leveraged by approximation algorithms, where early stopping and duality-gap control enable $(1+\epsilon)$ guarantees, and by dynamic algorithms that maintain relaxed LP solutions under updates.
}

\subsubsection{Flow-based algorithms} 

Similar to the flow-based algorithms for undirected graphs, the first flow-based DDS algorithm \cite{khuller2009finding} generally follows the same paradigm. As the DDS relates to two subsets $S^{*}$ and $T^{*}$, the algorithm first enumerates the possible ratio of $a=\frac{|S^{*}|}{|T^{*}|}$. For each ratio $a$, it guesses the density $g$ of the DDS via a binary search. According to the two parameters $a$ and $g$, the algorithm constructs a flow network based on the original directed graph. 
The flow network constructed based on the directed graph is shown in Fig. \ref{fig:di-flow-net}.
%
Then, the algorithm performs max-flow computation on the flow network and updates the binary search range based on the max-flow result. After all possible ratios $a$ are enumerated, the algorithm will output the maximum density across all binary searches, and the corresponding subgraph is the DDS. 



{\color{black}
The above flow-based algorithms provide a clean and rigorous way to encode the DDS objective and constraints, offering strong optimality guarantees.
However, their dependence on repeated max-flow computations and exhaustive enumeration of $O(n^2)$ ratio values severely limits scalability, making these methods practical only for small directed graphs.
}

To avoid expensive max-flow computations on the entire graph, Ma et al.~\cite{ma2020efficient} proposed the $[x,y]$-core for directed graphs, inspired by the $k$-core in undirected settings. 
Specifically, the $[x,y]$-core is defined as the maximal $(S,T)$-induced subgraph $G[S,T]$ such that every vertex in $S$ has outdegree at least $x$ and every vertex in $T$ has indegree at least $y$ within $G[S,T]$. The pair $[x,y]$ is referred to as the core number pair (cn-pair).



\begin{figure}[h]
    \centering
    \subfigure[{$[2,2]$}-core]{
        \label{fig:22core}
        \includegraphics[width=0.24\linewidth]{figures/core22.pdf}
    }
    \quad\quad
    \centering
    \subfigure[{$[1,2]$}-core]{
        \label{fig:12core}
        \includegraphics[width=0.23\linewidth]{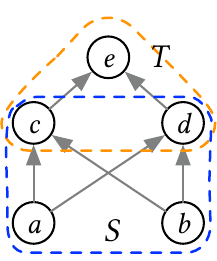}
    }
\caption{Examples of $[x,y]$-cores.}
\label{fig:xycores}
\end{figure}

Fig. \ref{fig:xycores} gives two examples of $[x,y]$-cores with different $[x,y]$ equals $[2,2]$ and $[1,2]$, respectively. The subgraph induced by $S=\{a,b\}$ and $T=\{c,d\}$ is the $[2,2]$-core. Ma et al. \cite{ma2020efficient} found the connection that the DDS can be located in some $[x,y]$-cores. Actually, the $[2,2]$-core is the DDS of the graph in Fig. \ref{fig:22core}. After locating the DDS in a $[x,y]$-core, the following max-flow computation can be performed on the flow network constructed based on this $[x,y]$-core, which is usually a small subgraph. Apart from reducing the time cost for each flow computation, they also proposed a divide-and-conquer strategy to reduce the number of the possible ratios $\frac{|S^{*}|}{|T^{*}|}$ by further exploiting the result of the max-flow computation. Later, Ma et al. \cite{ma2021directed} extended the $[x,y]$-core concept to weighted directed graphs and proposed efficient weighted DDS algorithms based on the weighted $[x,y]$-cores.

{\color{black}
{\bf Discussion.}
Exact algorithms establish the theoretical foundations of the DDS problem by precisely characterizing optimal solutions through LP and flow formulations.
While early methods rely on exhaustive enumeration of ratio parameters and repeated LP or max-flow computations, later refinements exploit intrinsic graph structure, such as $[x,y]$-cores, to substantially reduce practical cost.
Nevertheless, exact methods remain difficult to scale to large graphs.
These limitations motivate approximation algorithms, which relax LP and flow formulations to trade exact optimality for efficiency while preserving key structural insights revealed by exact solutions.
}

\subsection{Approximation algorithms}

The approximation DDS algorithms can also be categorized into different groups according to the main techniques used:
\begin{enumerate}
    \item peeling-based algorithms \cite{charikar2000greedy,khuller2009finding,bahmani2012densest};
    \item core-based algorithms \cite{ma2020efficient,luo2023scalable}
    \item linear/convex programming-based algorithm \cite{ma2022convex}
    \item max-flow-based algorithm \cite{chekuri2022densest}
\end{enumerate}

In the realm of approximation solutions:
For a 2-approximation ratio, Ma et al.~\cite{ma2020efficient} introduced a core-based algorithm with a worst-case time complexity of \(O(m\sqrt{m})\), while Luo et al.~\cite{luo2023scalable} proposed an algorithm with the same time complexity but better practical efficiency.
For approximation ratios less than 2, Ma et al.~\cite{ma2022convex} presented a convex programming-based algorithm named \texttt{CP-Approx} with a time complexity of \(O(\log_{1+\epsilon}n \cdot t_{\textsf{FW}})\).

\subsubsection{Peeling-based algorithms}

\begin{figure} 
    \centering
    \includegraphics[width=.6\linewidth]{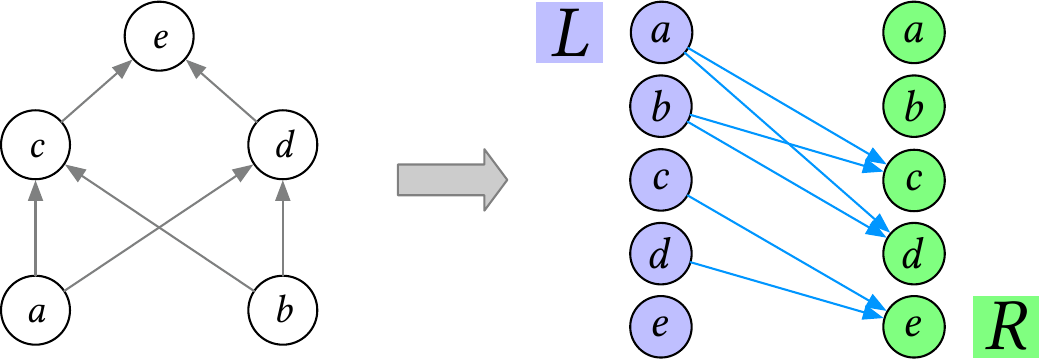}
    \caption{Duplicating vertices to two copies.}
    \label{fig:vertex-duplicate}
\end{figure}

\cite{charikar2000greedy} developed the first 2-approximation DDS algorithm based on peeling, {\tt BS-Approx}. Similar to the exact algorithms, {\tt BS-Approx} also enumerates all possible ratios of $a=\frac{|S^{*}|}{|T^{*}|}$. For each fixed ratio $a$, it duplicates the original vertices to two sets $L$ and $R$, as illustrated in Fig. \ref{fig:vertex-duplicate}. For each edge in the original graph, there is also a corresponding edge from $L$ to $R$. Next, if $\frac{|L|}{|R|} > a$, it peels a vertex with the lowest out-degree from $L$; otherwise, we peel a vertex with the lowest in-degree from $R$. The algorithm repeats the above peeling process until no more vertex exists in $L\cup R$. {\tt BS-Approx} keeps track of the subgraphs induced by $L$ and $R$ through the whole process for each $a$ and picks one with the maximum density as the output. We can observe that {\tt BS-Approx} has a quite high time complexity of $O(n^{2}\cdot (n+m))$ due to enumerating all $O(n^{2})$ possible ratios of $\frac{|S^{*}|}{|T^{*}|}$. 

To reduce the time cost of the 2-approximation algorithm, Khuller et al. \cite{khuller2009finding} proposed a different algorithm, {\tt KS-Approx}, which does not enumerate the ratios. Specifically, {\tt KS-Approx} first duplicates the vertices into two sets $L$ and $R$ as shown in Fig. \ref{fig:vertex-duplicate}. Next, it keeps peeling vertices with the smallest out-degree or in-degree from $L$ and $R$ until $L\cup R$ is empty. {\tt KS-Approx} with time complexity of $O(n+m)$ is much faster than {\tt BS-Approx} \cite{charikar2000greedy} as all vertices are only peeled once. However, the approximation ratio of {\tt KS-Approx} is larger than $2$ \cite{ma2021directed}. 
The improved {\tt FKS-Approx} achieves a 2-approximation with higher time complexity $O(n\cdot(n+m))$~\cite{ma2021directed}.

Apart from the approximation UDS algorithm, \cite{bahmani2012densest} also proposed an approximation DDS algorithm based on the MapReduce model, which can achieve an approximation ratio of $2\delta(1+\epsilon)$ where $\delta >1$ and $\epsilon>0$. Compared to the undirected version, the directed version peels vertices based on the guessed ratio of $c=\frac{|S^{*}|}{|T^{*}|}$: if $\frac{|L|}{|R|} \geq c$, it removes vertices from $L$ based on their out-degrees; otherwise, it removes vertices from $R$ based on their in-degrees. After enumerating all $O(\log{n}/\log{\delta})$ guesses of $c$, the algorithm will return the subgraph with the largest density. 
The approach is designed to efficiently solve the DDS problem in MapReduce and streaming environments.

\subsubsection{Core-based algorithms}

To derive a better 2-approximation algorithm, Ma et al. \cite{ma2020efficient} proposed a core-based algorithm, {\tt Core-Approx}. {\tt Core-Approx} is based on a key finding that the $[x^{*}, y^{*}]$-core is a 2-approximation solution to the DDS, where $[x^{*}, y^{*}]$ is the cn-pair with maximum $x*y$ among all $[x,y]$-cores. To find the $[x^{*},y^{*}]$-core efficiently, they first find the maximum equal cn-pair $[\gamma, \gamma]$ where $\gamma$ is the maximum $x$ value such that $[x,x]$-core exists. It is proven in \cite{ma2020efficient} $\gamma$ is asymptotically equal to $O(\sqrt{m})$. Next, {\tt Core-Approx} searches the largest $y$ for each fixed $x$ with $0\leq x \leq \gamma$, and searches the largest $x$ for each fixed $y$ with $0\leq x \leq \gamma$. 
The key cn-pair with the maximum $x\cdot y$ is the maximum cn-pair $[x^{*}, y^{*}]$. Then, we can obtain the $[x^{*}, y^{*}]$-core as the 2-approximation result via peeling. The overall time complexity of {\tt Core-Approx} is $O(\sqrt{m}(n+m))$ as $\gamma$ is bounded by $O(\sqrt{m})$.


To obtain DDS of large-scale directed graphs, Luo et al.~\cite{luo2023scalable} proposed a parallel approximation algorithm to efficiently compute $[x^{*},y^{*}]$-core.
Specifically, the authors first propose a subgraph model based on edge weights, namely $w$-core. For a $w$-core, the weight of each edge in the subgraph is not less than a given threshold $w$, and the weight of an edge is the product of the degrees of the vertices on both sides. Then the authors prove that $[x^{*},y^{*}]$-core is contained in the $w$-core with the largest weight, i.e., $w^*$-core. Therefore, a peeling-based approach is proposed to obtain the $w^*$-core of a directed graph and further obtain the $[x^{*},y^{*}]$-core as an approximate solution of the densest subgraph. Since the method only needs to decompose the graph once, the efficiency of the algorithm is greatly improved. The time cost of the algorithm is  $O(t\cdot m)$, where $t$ is bounded by the maximum out-degree/in-degree of all vertices in the directed graph.

\subsubsection{An LP-based algorithm}

To push for a better theoretical approximation ratio, Ma et al. \cite{ma2022convex} resort to LP-based techniques, as the duality gap between the primal and dual can be used to gauge the error. The $(1+\epsilon)$-approximation algorithm, {\tt CP-Approx}, given by Ma et al. \cite{ma2022convex}, shares the same algorithm framework as their exact algorithm. Both algorithms applied a divide-and-conquer strategy to reduce the number of LPs to be solved and used Frank-Wolfe iterations to optimize each LP. The difference is that the approximation algorithm can stop the Frank-Wolfe iterations earlier when the duality gap is within the given range required by $\epsilon$, while the exact algorithm needs to validate the exact solution via max-flow computations.

\subsubsection{A flow-based algorithm}

As discussed in UDS approximation algorithms, \cite{chekuri2022densest} designed an $(1+\epsilon)$-approximation algorithm for the UDS problem, which can also be extended to provide the $(1+\epsilon)$-approximation DDS in $\tilde{O}(\frac{m}{\epsilon^{2}})$ by $O(\frac{\log{n}}{\epsilon})$ calls to a $(1+\epsilon)$-approximation algorithm for the vertex-weighted UDS problem \cite{sawlani2020near}.

Apart from the above four categories, \cite{kannan1999analyzing} also proposed an $O(\log{n})$-approximation DDS algorithm based on randomized algorithms when they first introduced the DDS problem. 

{\color{black}
{\bf Discussion.}
Approximation algorithms form the practical core of DDS computation on large graphs.
Across peeling-, core-, LP-, and flow-based approaches, a common theme is the relaxation of exact formulations to achieve scalability while preserving key structural insights.
In particular, peeling and core-based methods emphasize locality and simplicity, which makes them especially amenable to parallel, streaming, and dynamic environments.
These properties naturally lead to the maintenance algorithms.
}

\subsubsection{DDS maintenance algorithms}

The existing DDS maintenance algorithms are based on $[x,y]$-cores \cite{ma2021directed} or linear programming \cite{sawlani2020near}. For the core-based algorithm, \cite{ma2021directed} developed algorithms to maintain the $[x^{*},y^{*}]$-core upon edge insertions and deletions, which can maintain the $2$-approximation result. 
Saurabh and Wang~\cite{sawlani2020near} proposed a fully dynamic UDS algorithm with an approximation of $1+\epsilon$ based on linear programming, as discussed in Section \ref{sec:outlineUndirected}. The algorithm can also be extended to maintain the $(1+\epsilon)$-approximation DDS on directed graphs by enumerating $O(\log_{1+\epsilon}{n})$ logarithmically spaced guesses from all possible ratios of $\frac{|S^{*}|}{|T^{*}|}$. For each specific ratio guess $c$, the algorithm constructs a vertex-weighted undirected graph based on the original directed graph and the given $c$ and maintains the approximation DS on the vertex-weighted graph.

{\color{black}
DDS maintenance algorithms extend static approximation techniques to dynamic and streaming settings by exploiting locality, sparsification, and amortized updates.
Rather than introducing fundamentally new paradigms, they reuse peeling, core, and LP-relaxation ideas as algorithmic building blocks.
This reuse underscores a recurring theme in DDS research: algorithmic concepts developed for static graphs often serve as robust cores for scalable solutions under evolving constraints.
}

\subsection{Variants of the original DDS problem}

There are mainly two variants of the original DDS problem. 
%
The first considers weighted directed graphs, where Ma et al.~\cite{ma2021directed} extend the $[x,y]$-core and show that it can still effectively locate the weighted DDS. The second introduces size constraints~\cite{khuller2009finding}, requiring $|S^{*}| \geq k_{1}$ and $|T^{*}| \geq k_{2}$. 
To solve this, \cite{khuller2009finding} enumerates ratios $a=\frac{i}{j}$ with $i\geq k_1$ and $j\geq k_2$, repeatedly extracts DDSs, removes their edges, and returns a feasible solution, achieving a $2$-approximation.

{\color{black}
These variants show that DDS naturally extends to weighted and size-constrained settings with the same ratio-based framework. While this preserves strong theoretical guarantees, it also inherits high computational cost, reinforcing the need for approximation and structure-aware methods on large graphs.
}

    \section{DSD on other types of graphs}
\label{sec:Others}

{\color{black}
This part reviews DSD variants on different graph models, including bipartite, multilayer, uncertain, heterogeneous (HINs), and hypergraphs.
Although density generalizes beyond simple graphs, it must be redefined to capture structural and semantic constraints such as cross-partite consistency, inter-layer dependencies, and probabilistic relations.
Despite these differences, most methods follow a common paradigm by extending density definitions or imposing structural constraints, resulting in trade-offs between expressiveness and efficiency.
Overall, these studies demonstrate the flexibility of the DSD framework while highlighting the need for more unified and scalable solutions.
}

\subsection{DSD on bipartite graphs}
We first introduce the definition of bipartite graphs.

\begin{definition}[\textbf{Bipartite graph}~\cite{wang2020efficient}]
	\label{def:bigraph}
	A bipartite graph is a graph with only two types (layers) of vertices, denoted by $\mathcal{B} = (V = (U, L), E)$, where $U$ is the set of vertices in the upper layer, $L$ is the set of vertices in the lower layer, $U \cap L = \varnothing$, and $E \subseteq U \times T$ denotes the edge set.
\end{definition}

Given a bipartite graph $\mathcal{B} = (V = (U, L), E)$, its density \cite{andersen2010local} is defined as $\rho(\mathcal{B}) = \frac{|E|}{\sqrt{|L||U|}}$.
Based on this definition, the DS in a bipartite graph is the subgraph with the largest bipartite graph density.
Given a query vertex $v$ and an integer $k$, Andersen~\cite{andersen2010local} proposed an algorithm for computing the bipartite DS containing $v$ with a size of $k$. The algorithm obtains the subgraph through local exploration since the result involves only a small part of the network.
Specifically, the algorithm generates a sequence of vectors $x_0,\cdots, x_T$ from an initial vector $x_0$. At each step, the vector is multiplied by the adjacency matrix $A$ of the bipartite graph, then normalized, and then pruned by zeroing each entry whose value is below the specified threshold. This pruning step reduces the number of non-zero elements, thus reducing the amount of computation required to compute the sequence. The author proved that the time complexity of the algorithm is $O(\Delta k^2)$, where $\Delta$ is the maximum degree of the vertex in the graph. In ~\cite{andersen2010local}, the author also proved that the density metric in bipartite graphs is equivalent to directed graphs, so the algorithms of a directed DS can also be used to compute the bipartite DS.

A more general case called $(p, q)$-biclique-based DS~\cite{mitzenmacher2015scalable,hooi2016fraudar} is proposed. 
In ~\cite{mitzenmacher2015scalable}, the authors proposed the concept of $(p, q)$-biclique density of bipartite graphs. A $(p, q)$-biclique is a biclique with exactly $p$ and $q$ vertices on the upper and lower layers, respectively.

\begin{definition}[\textbf{$(p,q)$-biclique density}~\cite{mitzenmacher2015scalable}]
	\label{def:bicliquedensity}
	Given two integers $p$ and $q$, the $(p, q)$-biclique density of a bipartite graph $\mathcal{B}$ is 
$\rho_{p,q}(\mathcal{B}) = \frac{c_{p,q}(\mathcal{B})}{|V|}$,
where $c_{p,q}(\mathcal{B})$ is the number of $(p,q)$-bicliques in $\mathcal{B}$.
\end{definition}

Based on Definition \ref{def:bicliquedensity}, the $(p, q)$-biclique DS is the subgraph with the largest $(p, q)$-biclique density among all subgraphs in $\mathcal{B}$, where $p, q \leq 1$.
Mitzenmacher et al.~\cite{mitzenmacher2015scalable} convert the $(p, q)$-biclique DS problem into a decision problem; that is, whether there is a subgraph whose $(p, q)$-biclique density is not less than $D$ in the bipartite graph. Then the DS can be obtained by binary search. For the decision problem, the bipartite graph is transformed into a flow network, which is subsequently solved by computing its min-cut. In addition, the authors also propose a sampling-based approximation algorithm to handle large-scale bipartite graphs.

\subsection{DSD on multilayer graphs}
We first introduce the definition of multilayer graphs.

\begin{definition}[\textbf{Multilayer graph}~\cite{jethava2015finding,galimberti2017core,galimberti2020core}]
	\label{def:multilayer}
	A multilayer graph is denoted by $\mathcal{H}=(V$, $E=(E_1$, $E_2$, $\cdots$, $E_l))$, where $V$ is the set of vertices with the same type, $E_i (i \in [1, l])$ is the set of edges with the $i$-th edge type, and $l$ is the number of layers or edge types.
\end{definition}

The density definition on multilayer graphs, namely common density~\cite{jethava2015finding}, is extended from edge-density.

\begin{definition}[\textbf{Common density}~\cite{jethava2015finding}]
    \label{def:common-density}
    Given a multilayer graph $\mathcal{H}=(V, E=(E_1, E_2, \cdots, E_l))$ and a vertex set $S$ of $\mathcal{H}$, the common density of $S$ is $\rho(\mathcal{H},S) = \min\limits_{i \in \{1, \cdots, l\}}\rho(G_i,S) = \min\limits_{i \in \{1, \cdots, l\}} \frac{|E_i(S)|}{|S|}$,
    where $E_i(S)$ is the number of edges connecting vertices of $S$ in the $i$-th layer graph of $\mathcal{H}$.
\end{definition}

According to Definition~\ref{def:common-density}, the common DS of a multilayer graph $\mathcal{H}$ is the vertex subset maximizing the minimum edge density across all layers. Jethava et al.~\cite{jethava2015finding} propose a greedy algorithm that iteratively removes vertices from the layer with the smallest density, but it lacks theoretical approximation guarantees.

A key limitation of common density is that it treats all layers equally, allowing insignificant layers to dominate the result. To address this, Galimberti et al.~\cite{galimberti2017core,galimberti2020core} introduce multilayer density, which balances density and the number of supporting layers via a parameter $\beta$ (smaller $\beta$ emphasizes density). Notably, when $\beta=0$ and all layers are considered, it reduces to the common density.



\begin{definition}[\textbf{Multilayer density}~\cite{galimberti2017core}]
	\label{def:multi-density}
	Given a multilayer graph $\mathcal{H}{=}(V, E{=}(E_1, E_2, \cdots, E_l))$ and a a positive real number $\beta$, the multilayer graph density of the vertex set $S$ of $\mathcal{H}$ is 
    $\delta(\mathcal{H},S) = \max\limits_{\hat{L} \subseteq \{1,\cdots,l\}} \min\limits_{i\in \hat{L}} \frac{|E_i(S)|}{|S|}|\hat{L}|^{\beta}$.
\end{definition}


\begin{definition}[\textbf{Multilayer ${\bf k}$-core}~\cite{galimberti2017core}]
	\label{def:multi-core}
	Given a multilayer graph $\mathcal{H}{=}(V, E{=}(E_1, E_2, \cdots, E_l))$ and a one-dimensional vector $k=(k_1,\cdots,k_l)$, the multilayer $k$-core of $\mathcal{H}$ is a multilayer subgraph $\mathcal{H}'=(V', E'{=}(E'_1, E'_2, \cdots, E'_l))$ of $\mathcal{H}$, satisfying the induced subgraph of $V'$ is a $k_i$-core in the $i$-th layer graph of $\mathcal{H}'$.
\end{definition}

To compute multilayer DS, they define multilayer $k$-core, where a vertex set forms a $k_i$-core in each layer, and then enumerate all such cores via multilayer $k$-core decomposition~\cite{galimberti2017core,galimberti2020core}. The solution maximizing multilayer density is returned, with an approximation ratio of $\frac{1}{2l^{\beta}}$.




\subsection{DSD on uncertain graphs}
We first introduce the uncertain graph model.

\begin{definition}[\textbf{Uncertain graph}~\cite{zou2013polynomial}]
	\label{def:uncertain}
	An uncertain graph is a graph $\mathcal{U} = (V,E,P)$, where $V$ is the vertex set, $E$ is the edge set, and $P$ is a function associated with each edge $e \in E$ with an existence probability $P(e) \in (0, 1]$.
\end{definition}


The probability that an uncertain graph $\mathcal{G}=(V,E,P)$ realizes an exact graph $G=(V,E')$ is 
$Pr[\mathcal{G} \Rightarrow G] = \prod_{e\in E'} P(e)\prod_{e\in E \setminus E'} (1-P(e))$.
Zou et al.~\cite{zou2013polynomial} define the expected density as 
$\bar{\rho}(\mathcal{G}) = \sum_{G \subseteq \Omega(\mathcal{G})} \rho(G)\, Pr[\mathcal{G} \Rightarrow G]$, 
where $\Omega(\mathcal{G})$ is the set of all possible realizations. 
Thus, $\bar{\rho}(\mathcal{G})$ is the expected density of a randomly realized graph, and the uncertain DS maximizes this value.
They further show that for any subgraph $\mathcal{G}'=(V',E',P)$, $\bar{\rho}(\mathcal{G}') = \sum_{e \in E'} \frac{P(e)}{|V'|}$. 
Treating $P(e)$ as edge weights reduces the problem to a weighted densest subgraph, which can be solved exactly via max-flow~\cite{goldberg1984finding}.

Miyauchi et al.~\cite{miyauchi2018robust} studied the DS problem with uncertain edge weights in uncertain graphs. Given an edge-weight space $W$, which contains unknown edge-weight vectors, $W$ can be considered as the product of the confidence intervals of the true edge weights, each of which can be obtained in practice from theoretically guaranteed lower and upper bounds or repeated sampling of the true edge weight estimates. Given a subgraphs $S$ of the uncertain edge weights graph $G$, its robust ratio is $\min\limits_{w \in W} \frac{f_w(S)}{f_w(S_w^*)}$,
where $f_w(S)$ is the weighted edge density of $S$, and $S_w^*$ is the DS of $G$ under edge weight vector $w$.
The robust DS is the subgraph in $G$ that maximizes the robust ratio, and it can be computed by a sampling oracle algorithm based on robust optimization with theoretical guarantees.

Tsourakakis et al.~\cite{tsourakakis2019novel} explored the risk-averse DSD problem in uncertain graphs. 
In this context, they aim to find a set of vertices $S$ in an uncertain graph $\mathcal{G}$ that induces a dense subgraph with high density and low average associated risk, where the average associated risk is the average variance of edges in the subgraph. 
To solve this problem, they formulated it as a DSD problem in a graph with both positive and negative edge weights and applied a peeling-based algorithm.

\subsection{DSD on heterogeneous information networks}
We first introduce the definition of HINs.

\begin{definition}[\textbf{HIN}~\cite{sun2011pathsim}]
    \label{def:hin}
    An HIN is a directed graph $\mathcal{H}=(V, E)$ with a vertex type mapping function $\psi : V \rightarrow \mathcal{A}$ and an edge type mapping function $\phi : E \rightarrow \mathcal{R}$, where each vertex $v \in V$ belongs to a vertex type $\psi(v) \in \mathcal{A}$, and each edge $e \in {E}$ belongs to an edge type (also called relation) $\phi(e) \in \mathcal{R}$, and $|\mathcal{A}| +|\mathcal{R}| > 2$.
\end{definition}

The network schema abstracts an HIN through vertex and edge types mapping. A meta-path $\mathcal{P}$ on the schema is represented as $A_1 \xrightarrow{R_1} A_2 \xrightarrow{R_2} \cdots \xrightarrow{R_{i-1}} A_i$, defining a composite relation between $A_1$ and $A_i$. For simplicity, $\mathcal{P}$ is written as $(A_1, A_2, \ldots, A_i)$.
Given a meta-path $\mathcal{P} = (A_1, A_2, \ldots, A_i)$, the $\mathcal{P}$-family is defined as a vertex set family $\mathcal{V} = \{V_1, \ldots, V_i\}$, where for $1 \leq j \leq i$, $V_j \subseteq V(A_j)$ and $|V_j| \neq \varnothing$. The $\mathcal{P}$-family induced subgraph, $G(\mathcal{V})$, refers to the multipartite subgraph induced by $\{V_1 \cup \cdots \cup V_i\}$.
Building on these concepts, Chen et al.~\cite{chen2023densest} introduced the notion of multipartite density for HINs, formally defined as follows:


\begin{definition}[\textbf{Multipartite Density}~\cite{chen2023densest}]
\label{def:hin-density}  
Given a $\mathcal{P}$-family of vertex sets $\mathcal{V} = \{V_1, \ldots, V_i\}$, let $\mathcal{H}(\mathcal{V}) = V_1 \times \cdots \times V_i$. The multipartite density is defined as $\rho = \frac{|\mathcal{F}(\mathcal{V})|}{|\mathcal{H}(\mathcal{V})|^{1/i}}$, where $\mathcal{F}(\mathcal{V})$ denotes the set of $\mathcal{P}$-instances in $G(\mathcal{V})$.
\end{definition}

The HIN DS problem seeks to identify the $\mathcal{P}$-family that maximizes the multipartite density for a given meta-path $\mathcal{P}$.  
To address this problem, Chen et al.~\cite{chen2023densest} proposed an exact algorithm by introducing a novel set of parameters, $\mathcal{M} = \{m_1, \dots, m_i\}$, within the fractional programming framework. 
Specifically, the algorithm systematically explores all possible configurations of $\mathcal{M}$ and identifies the maximum density by iteratively computing the density using a max-flow approach.  
The time complexity of this exact algorithm is $\Theta(|\mathbb{M}||\mathbb{F}|)$, where both $|\mathbb{M}|$ and $|\mathbb{F}|$ are bounded by $O((\frac{n}{i})^i)$.  
Additionally, the authors developed an approximation algorithm that achieves an approximation ratio of $i$. This algorithm greedily removes vertices for each $\mathcal{M}$. 
The time complexity for each $\mathcal{M}$ is $O\left(i|P| + i|V_{\text{max}}|\log(|V_{\text{max}}|)\right)$, where $|P|$ represents the number of $\mathcal{P}$ instances in $G(\mathcal{V})$, and $|V_{\text{max}}| = \max\{|V_1|, \dots, |V_i|\}$. 
To further improve the efficiency of the algorithm, the authors proposed several pruning strategies to reduce the number of possible configurations of $\mathcal{M}$ and the number of vertices to be processed. After these improvements, the time complexity of the exact algorithm is $O\left((\frac{n}{i})^i |\mathbb{F}|\right)$.

\textbf{Impact of meta-path selection.} Unlike homogeneous DSD, the density and semantics of an HIN subgraph are conditioned on the query meta-path $\mathcal{P}$. Different meta-paths represent different semantic relations and induce different sets of instances $\mathcal{F}_{\mathcal{P}}$, and can therefore produce substantially different dense subgraphs. Meta-path length also affects efficiency: increasing $i=|\mathcal{P}|$ enlarges the number of vertex types and potential path instances, while the enumeration-based exact algorithm has an $O((n/i)^i)$ dependence on $i$. Hence, selecting $\mathcal{P}$ involves both semantic-quality and computational trade-offs rather than being merely a preprocessing choice. 

\textbf{Beyond meta-path density.} Meta-graphs and heterogeneous motifs can express branching or other higher-order typed patterns that cannot be represented by a single meta-path. Such structures have been studied extensively in other HIN mining tasks, but explicit densest-subgraph objectives based on meta-graphs or general heterogeneous patterns remain comparatively underexplored. Extending multipartite density from path instances to richer heterogeneous pattern instances is therefore a promising direction for HIN DSD.

Higher-order typed patterns, such as meta-structures and heterogeneous motifs \cite{huang2016meta,rossi2019heterogeneous,zhou2024efficient}, have been extensively studied in HIN mining and cohesive-subgraph discovery. However, their explicit incorporation into density-maximization objectives remains comparatively underexplored.

\textbf{Discussion.}
For HIN DSD, both algorithm choice and result quality depend strongly on the selected meta-path.
Exact flow-based methods are suitable when high accuracy is required and
the number of meta-path instances is manageable, whereas peeling-based
approximations are more scalable for large HINs.
Short and semantically focused meta-paths generally incur lower
computational cost, while longer or richer patterns can capture more
specific semantics but may substantially enlarge the search space.
Therefore, practical HIN DSD requires balancing semantic expressiveness,
solution quality, and computational cost.

\subsection{DSD on hypergraphs}

{\color{black}

In many real-world applications, objects are connected not only by pairwise relations but also by higher-order interactions involving more than two entities. Such relations cannot be adequately represented by standard graphs with edges. This naturally motivates the study of the densest subgraph problem in hypergraphs.

\begin{definition}[\textbf{Hypergraph}~\cite{arafat2023neighborhood}]
A hypergraph $G = (V , E)$ consists of a set of vertices $V$
and a set of hyperedges $E \subseteq P(V)$, where $P(V)$ is the power set of $V$. A hyperedge is an unordered set of vertices.
\end{definition}

In a hypergraph $ G = (V, E) $, the neighbors of a vertex $ v $ are the vertices that co-occur with $ v $ in some hyperedge. The rank of a hyperedge is its cardinality, and the rank of the hypergraph, denoted by $r$, is the maximum hyperedge size.

\textbf{Overview.}
Hypergraph DSD extends classical DSD by allowing density to capture
higher-order interactions.
Existing formulations differ mainly in how the contribution of a
hyperedge or its induced neighborhood structure is measured, ranging
from unweighted and weighted hyperedge density to neighborhood-based
and generalized supermodular objectives.
Table~\ref{tab:hyper_density} summarizes representative density
formulations, while Table~\ref{tab:hypergraph_dsd} compares their
algorithmic paradigms and rank dependence.

\textbf{Hyperedge-density-based models.}
The basic formulation measures the number of induced hyperedges per
vertex,
$\rho(S)=|E(S)|/|S|$.
Its weighted extension,
$\rho_w(S)=\sum_{e\in E(S)}w(e)/|S|$,
allows different hyperedges to represent interactions with different
strengths or frequencies.
Hu et al.~\cite{hu2017maintaining} developed flow- and LP-based exact
algorithms, together with an $r$-approximate peeling algorithm and
dynamic approximation algorithms whose guarantees depend on the
hypergraph rank.
Bera et al.~\cite{bera2022new} further studied weighted dynamic
hypergraphs and achieved a $(1+\epsilon)$ approximation whose ratio is
independent of $r$.

\textbf{Beyond hyperedge density.}
Alternative formulations capture richer structural information than
induced hyperedge counts.
Arafat et al.~\cite{arafat2023neighborhood} introduced
\emph{volume-density},
\begin{equation}
    \rho(S)=\frac{\sum_{u\in S}|N_S(u)|}{|S|},
\end{equation}
which measures the pairwise neighborhood structure induced by
higher-order hyperedges.
They developed a peeling-based algorithm with approximation ratio
$d_{\rm pair}(d_{\rm card}-2)+2$ and time complexity
$O(|V|d_{\rm nbr}(d_{\rm nbr}+d_{\rm hpe}))$.
More generally, supermodular formulations
~\cite{chekuri2022densest,HuangGV24} replace the hyperedge-count
objective with a nonnegative supermodular set function $f(\cdot)$, providing a general framework for weighted and structure-aware density measures.
However, explicit density objectives defined over particular
higher-order motifs or subhypergraph patterns remain comparatively underexplored.

\textbf{Variants of hypergraph DSD.}
Another important direction incorporates locality constraints. 
Huang et al.~\cite{HuangGV24} proposed the anchored densest subhypergraph model, which enforces a seed set and optimizes a regularized density objective. 
They designed both global and strongly-local algorithms: the global method runs in $O(\mathrm{poly}(|V|,|E|))$ time via flow-based optimization, while the local algorithm runs in $O(|V_S|+|E_S|)$ time, depending only on the explored subhypergraph. 
This establishes a clear trade-off between scalability and global optimality.

Balalau et al. \cite{balalau2024finding} introduce the $(k, \alpha)$-DSLO problem for identifying up to $k$ dense subhypergraphs under a weighted Jaccard overlap constraint, prove its NP-hardness even for $\alpha = 0$, and propose an LP-based framework that computes minimal densest subhypergraphs with only $O(\log |V|)$ solver calls.

\textbf{Impact of hypergraph rank.}
The hypergraph rank $r$ can affect both approximation quality and
computational cost.
For Hu et al.~\cite{hu2017maintaining}, the static peeling algorithm
has an $r$-approximation, while their dynamic guarantees also explicitly
depend on $r$.
Bera et al.~\cite{bera2022new} remove this dependence from the
approximation ratio, although the update complexity still depends on
the rank.
For volume-density, the approximation factor depends on the maximum
hyperedge cardinality $d_{\rm card}$, which is bounded by $r$.
Hence, large hyperedges may weaken approximation guarantees and increase
the cost of processing local hypergraph structures, depending on the algorithmic framework.

\textbf{Algorithmic frameworks.}
Existing approaches mainly include
(i) flow- or LP-based exact algorithms,
(ii) peeling-based approximation algorithms,
and (iii) dynamic maintenance methods.
Compared with ordinary graphs, hypergraph algorithms must additionally
handle the potentially large number of vertices affected by each
hyperedge, making rank and neighborhood complexity important factors in
both theoretical and practical performance.

\textbf{Summary.}
Hypergraph DSD therefore extends beyond simple hyperedge counting to
weighted interactions, neighborhood-based density, generalized
set-function objectives, and constrained variants.
The choice of density formulation affects not only the structural
semantics of the returned subhypergraph but also the approximation
guarantee and rank dependence.
In practice, flow- and LP-based methods are preferable when exact
solutions are required on moderate-sized hypergraphs, whereas
peeling-based methods are more suitable for large-scale analysis due to
their lower computational overhead.
Dynamic methods are attractive for evolving hypergraphs, although their
update cost may increase with the hypergraph rank.
Developing density measures that explicitly capture recurring
higher-order subgraph patterns remains a promising direction.

\begin{table*}[t]
\centering
\caption{{\color{black}Comparison of representative density formulations
for hypergraph DSD.}}
\label{tab:hyper_density}
\setlength{\tabcolsep}{6pt}
\renewcommand{\arraystretch}{1.15}
\begin{tabular}{c|c|c}
\hline
\textbf{Density} & \textbf{Definition} & \textbf{Structural Semantics} \\
\hline
\hline

Hyperedge density
& $\displaystyle \rho(S)=|E(S)|/|S|$
& Number of induced higher-order interactions \\

Weighted hyperedge density
& $\rho_w(S)=\sum_{e\in E(S)}w(e)/|S|$
& Strength/frequency of induced interactions \\

Volume density
& $ \rho_N(S)=
\sum_{u\in S}|N_S(u)|/|S|$
& Neighborhood/co-occurrence structure induced by hyperedges \\

Supermodular density
& $\displaystyle \rho_f(S)=f(S)/|S|$
& General weighted or structure-aware contributions \\

\hline
\end{tabular}
\end{table*}

\begin{table*}[t]
\centering
\caption{{\color{black}Comparison of representative hypergraph DSD
methods. ``--'' indicates that the corresponding dependence or
guarantee is not explicitly characterized in the original work.}}
\label{tab:hypergraph_dsd}
\setlength{\tabcolsep}{4pt}
\renewcommand{\arraystretch}{1.15}
\begin{tabular}{c|c|c|c|c|c}
\hline
\textbf{Work}
& \textbf{Formulation}
& \textbf{Method}
& \textbf{Guarantee}
& \textbf{Approx. Rank Dep.}
& \textbf{Time Rank Dep.} \\
\hline
\hline

Hu et al.~\cite{hu2017maintaining}
& Hyperedge
& Exact / Peeling / Dynamic
& $r$; $r(1+\epsilon)$; $r^2(1+\epsilon)$
& Yes
& Yes \\

Bera et al.~\cite{bera2022new}
& Weighted hyperedge
& Dynamic / Orientation
& $1+\epsilon$
& No
& Yes \\

Arafat et al.~\cite{arafat2023neighborhood}
& Volume
& Peeling
& $d_{\rm pair}(d_{\rm card}-2)+2$
& Yes
& Structural params. \\

Huang et al.~\cite{HuangGV24}
& Supermodular / Anchored
& Exact / Strongly local
& Exact
& N/A
& -- \\

Balalau et al.~\cite{balalau2024finding}
& Multiple / Overlap-constrained
& LP-based
& --
& --
& -- \\

\hline
\end{tabular}
\end{table*}
    \section{Comparison analysis}
\label{sec:comparison}


In this section, we analyze the relationships among different density definitions and compare DSD solutions.

\subsection{Comparison of edge-density extensions across graph types}


DSD originates from undirected graphs, where density is defined as $|E|/|V|$. It has been extended to various graph types, including directed, bipartite, multi-layer, uncertain, HINs, and hypergraphs to capture their structural characteristics. 
Table~\ref{tab:sum-density} summarizes these definitions, and Fig.~\ref{fig:density} illustrates their relationships.

\begin{table*}[h]
    \centering
    \captionof{table}{Summary of density definitions.}
    \label{tab:sum-density}
        \begin{tabular}{c|c|c}
    \hline
    Graph type & Category & Definition \\
    \hline
    \hline
    \multirow{3}[0]{*}{Undirected graphs} & Undirected edge-density~\cite{goldberg1984finding}  & $\rho(G)= \frac{{|E|}}{{|V|}}$ \\
    \cline{2-3}
          & Clique-density~\cite{tsourakakis2015k,mitzenmacher2015scalable}  &  $\rho(G, \Psi) = \frac{{u(G, \Psi)}}{{|V|}}$ \\
    \cline{2-3}
          & Pattern-density \cite{fang2019efficient}  &  $\rho(G, \Psi) = \frac{{u(G, \Psi)}}{{|V|}}$ \\
    \hline
    Directed graphs & Directed edge-density~\cite{kannan1999analyzing} & $\rho(S, T)= \frac{{|E(S, T)|}}{{\sqrt{|S||T|}}}$ \\
    \hline
    \multirow{2}[0]{*}{Bipartite graphs} & Bipartite edge-density~\cite{andersen2010local}  & $\rho(\mathcal{B}) = \frac{|E|}{\sqrt{|L||U|}}$ \\
    \cline{2-3}
          &  $(p,q)$-biclique density~\cite{mitzenmacher2015scalable,hooi2016fraudar}  & $\rho_{p,q}(\mathcal{B}) = \frac{c_{p,q}(\mathcal{B})}{|V|}$ \\
    \hline
    \multirow{2}[0]{*}{Multi-layer graphs} & Common density~\cite{jethava2015finding}  & $\rho(\mathcal{H},S) = \min\limits_{i \in \{1, \cdots, l\}} \frac{|E_i(S)|}{|S|}$ \\
    \cline{2-3}
          &  Multi-layer density~\cite{galimberti2017core}  & $\delta(\mathcal{H},S) = \max\limits_{\hat{L} \subseteq \{1,\cdots,l\}} \min\limits_{i\in \hat{L}} \frac{|E_i(S)|}{|S|}|\hat{L}|^{\beta}$ \\
    \hline
    \multirow{2}[0]{*}{Uncertain graphs} & Expected density~\cite{zou2013polynomial}  & $\bar{\rho}(\mathcal{G})=\sum\limits_{G\subseteq \Omega(\mathcal{G})} \rho(G)Pr[\mathcal{G} {\Rightarrow} G]$ \\
    \cline{2-3}
          &  Robust ratio~\cite{miyauchi2018robust}  & $\min\limits_{w \in W} \frac{f_w(S)}{f_w(S_w^*)}$ \\
    \hline
    HINs & Multipartite density~\cite{chen2023densest} & $\rho = \frac{|\mathcal{F}(\mathcal{V})|}{|\mathcal{H}(\mathcal{V})|^{\frac{1}{i}}}$ \\
    \hline
    \multirow{2}[0]{*}{Hypergraphs} & Hyper edge density~\cite{hu2017maintaining}  & $\rho(G)= \frac{{|E|}}{{|V|}}$ \\
    \cline{2-3}
          &  Volume density~\cite{arafat2023neighborhood}  & $\rho[S] = \frac{\sum_{u\in S}|N_S(u)|}{|S|}$ \\
    \hline
    \end{tabular}%
\end{table*}

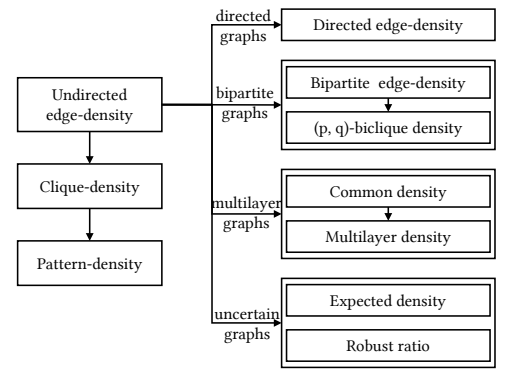
\begin{figure}[h]
\centering
\resizebox{\linewidth}{!}{
\begin{tikzpicture}[
    box/.style={
        draw,
        rounded corners=2pt,
        align=center,
        font=\small,
        text width=28mm,
        minimum height=7mm,
        line width=0.5pt
    },
    arrow/.style={
        ->, 
        >=Stealth, 
        thin,
        rounded corners=2pt
    },
    title/.style={
        font=\bfseries\small,
        anchor=west
    }
]

\node[title] at (-2, 0.8) {(a) Variants};
\node[box] (a1) {Edge density};
\node[box, right=10mm of a1] (a2) {Clique density};
\node[box, right=10mm of a2] (a3) {Pattern density};

\draw[arrow] (a1) -- (a2);
\draw[arrow] (a2) -- (a3);

\node[title] at (-2, -1.2) {(b) Extensions};

\node[box, below=2.5cm of a1] (root) {Undirected edge-density};

\node[box, right=15mm of root] (m) {Common density};
\node[box, above=3mm of m] (b) {Bipartite edge-density};
\node[box, above=3mm of b] (d) {Directed edge-density};

\node[box, below=3mm of m] (u) {Expected density};
\node[box, below=3mm of u] (h) {Multipartite density};
\node[box, below=3mm of h] (hy) {Hyper edge-density};

\node[box, right=12mm of b] (b2) {$(p,q)$-biclique density};
\node[box, right=12mm of m] (m2) {Multilayer density};
\node[box, right=4mm of u] (u2) {Robust ratio};
\node[box, right=12mm of hy] (hy2) {Volume density};

\foreach \x in {d,b,m,u,h,hy}
    \draw[arrow] (root.east) -- ++(0.6, 0) |- (\x.west);

\draw[arrow] (b) -- (b2);
\draw[arrow] (m) -- (m2);
\draw[arrow] (u)  (u2);
\draw[arrow] (hy) -- (hy2);

\end{tikzpicture}
}
\captionof{figure}{Relationship among various density definitions.}
\label{fig:density}
\end{figure}

For directed graphs, density is defined as $\rho=|E|/\sqrt{|S||T|}$, which reduces to undirected density when $|S|=|T|$. Bipartite density follows a similar form, and $(p,q)$-biclique density further generalizes it by counting bicliques instead of edges.
In multi-layer graphs, common density is defined as the minimum density across layers, while multilayer density balances density and the number of supporting layers. For uncertain graphs, expected density replaces edge counts with probabilities, and robust ratio evaluates stability under uncertainty.
For HINs, density is defined over induced multipartite graphs based on meta-paths. In hypergraphs, edge density extends naturally, while volume density replaces edge counts with aggregated neighborhood size.
Additionally, edge counts in undirected graphs can be replaced by higher-order structures such as cliques or patterns.

\subsection{Comparison of DSD solutions}

Most existing studies focus on the original DSD problems on undirected and directed graphs. We therefore compare solutions for the UDS and DDS problems separately.

\subsubsection{Solutions for original UDS problem}
Tables \ref{tab:sum-exact-uds} and \ref{tab:sum-app-uds} summarize the exact and approximate algorithms for undirected graphs, respectively.

\begin{table}[h]
  \centering
  \caption{Summary of exact UDS algorithms, $h\leq n^2$. $t_{\textsf{Flow}}$ denotes the time cost of the max-flow algorithm.}
    \begin{tabular}{c|l|l}
    \hline
    Category & Algorithm & Time Complexity \\
    \hline
    \hline
    \multirow{2}[0]{*}{Flow-based} & {\tt Exact} \cite{goldberg1984finding}  & $O(\log n\cdot t_{\textsf{Flow}})$ \\
    \cline{2-3}
          & {\tt CoreExact} \cite{fang2019efficient}  & $O(\log n\cdot t_{\textsf{Flow}})$ \\
    \hline
    \multirow{2}[0]{*}{LP-based} & {\tt LP-Exact} \cite{charikar2000greedy}  & $\Omega(n^4)$ \\
    \cline{2-3}
          & {\tt CP-Exact} \cite{danisch2017large}  & $O(h\cdot m+\log h\cdot t_{\textsf{Flow}})$ \\
    \hline
    \end{tabular}%

    \small
  \label{tab:sum-exact-uds}%
\end{table}%

\begin{table*}
  \centering
  \caption{Summary of approximation UDS algorithms, $\epsilon>0$ is a real value; $\rho^*$ and $\Delta(G)$ denote the maximum density and maximum degree of $G$, respectively.}
    \begin{tabular}{c|l|c|l}
    \hline
    Category & Algorithm & Approx. ratio & Time complexity \\
    \hline
    \hline
    \multirow{4}[0]{*}{Peeling-based} & {\tt Greedy} \cite{charikar2000greedy} & 2 & $O(m+n)$ \\
    \cline{2-4}
          & {\tt BatchPeel} \cite{bahmani2012densest} & 2(1+$\epsilon$) & $O(\frac{m\log n}{\epsilon})$ \\
    \cline{2-4}
          & {\tt Greedy++} \cite{boob2020flowless} & 1+$\epsilon$ & $O(m\log n \cdot \frac{\Delta(G)\log m}{\rho^* \epsilon^2})$ \\
    \hline
          Core-based & {\tt CoreApp} \cite{fang2019efficient} & 2 & $O(m+n)$ \\
    \hline
    \multirow{4}[0]{*}{LP-based} & {\tt MWU} \cite{bahmani2014efficient} & 1+$\epsilon$ & $\widetilde{O}(\frac{m\log n}{\epsilon^2})$ \\
    \cline{2-4}
          & {\tt Width-Approx} \cite{boob2019faster} & 1+$\epsilon$ & $O(\log n \cdot \frac{m\Delta(G)}{\epsilon})$ \\
    \cline{2-4}
          & {\tt Frank-Wolfe} \cite{danisch2017large} & 1+$\epsilon$ & $O(m \cdot \frac{m\Delta(G)}{\epsilon^2})$ \\
    \cline{2-4}
          & {\tt FISTA} \cite{harb2022faster} & 1+$\epsilon$ & $O(m \cdot \frac{\sqrt{m \Delta(G)}}{\epsilon})$ \\
    \hline
    Flow-based  & {\tt Flow-Approx} \cite{chekuri2022densest} & 1+$\epsilon$ & $O(m\cdot \frac{\log^2 m}{\epsilon})$ \\
    \hline
    \end{tabular}%

  \label{tab:sum-app-uds}%
\end{table*}

Exact algorithms fall into two categories: max-flow-based~\cite{goldberg1984finding,fang2019efficient} and LP-based methods~\cite{charikar2000greedy,danisch2017large}.
The classical flow-based algorithm~\cite{goldberg1984finding} computes the maximum density via binary search, where each step requires solving a min-cut problem, resulting in high computational cost. Fang et al.~\cite{fang2019efficient} improve efficiency by restricting computation to a $k$-core, thereby reducing the flow network size.
LP-based methods are also expensive. To address this, the dual of a relaxed LP is solved using the Frank-Wolfe method~\cite{danisch2017large}, followed by a flow-based procedure to recover the exact solution.
{\color{black}
Overall, existing studies indicate that exact algorithms (e.g., {\tt Exact}) are effective for medium-scale graphs, where optimal solutions can be obtained within a reasonable time, whereas approximation algorithms are generally preferred for large-scale graphs due to their superior scalability.
}


Approximation algorithms for the UDS problem can be broadly categorized into four classes: peeling-based methods~\cite{charikar2000greedy,bahmani2012densest,boob2020flowless}, core-based methods~\cite{fang2019efficient}, LP-based methods~\cite{bahmani2014efficient,danisch2017large,su2020distributed,harb2022faster}, and flow-based methods~\cite{chekuri2022densest}. These approaches differ fundamentally in how they trade off approximation quality, computational cost, and the extent to which they exploit graph structure.
Peeling-based algorithms adopt the peeling paradigm and achieve linear-time complexity with a factor-2 approximation, while refined variants trade additional iterations or bookkeeping for improved accuracy~\cite{charikar2000greedy,bahmani2012densest,boob2020flowless}.
Core-based methods can be seen as a structure-aware special case of greedy peeling that directly returns the $k_{\max}$-core, retaining linear-time efficiency but without approximation guarantees better than 2~\cite{fang2019efficient}.
LP-based algorithms relax the original formulation and solve its dual iteratively, providing $(1+\epsilon)$-approximation guarantees at the cost of rapidly increasing running time as $\epsilon$ decreases~\cite{bahmani2014efficient,danisch2017large,harb2022faster}.
Flow-based approximation algorithms rely on partial max-flow computations to estimate density, offering strong theoretical guarantees but incurring higher overhead due to repeated blocking-flow procedures~\cite{chekuri2022densest}.

\textbf{Practical comparison.} Theoretical complexity alone does not fully predict practical performance, since graph size, sparsity, reduction effectiveness, and per-iteration overhead can substantially affect runtime. The large-scale empirical study in~\cite{zhou2024depth} confirms this gap across real graphs from biological, collaboration, social, Web, and other domains. For example, although {\tt FISTA} has a better theoretical iteration bound than several other convex-programming methods, its projection step can make it slower in practice; conversely, simple peeling-based methods such as {\tt Greedy++} often reach high-quality solutions with far fewer iterations than their worst-case analysis suggests. Graph reduction is particularly effective on large graphs, pruning more than $98\%$ of edges in some cases and yielding speedups of up to three orders of magnitude. The approximation parameter also induces a clear accuracy--efficiency trade-off: decreasing $\epsilon$ provides a tighter guarantee but generally requires more iterations and running time. 
Similarly, increasing the iteration budget $T$ may improve solution quality but also increases runtime, while empirical results show that high-quality solutions often emerge well before the worst-case iteration bound.
Thus, peeling- and core-based methods are preferable when scalability and low overhead are the primary concerns, whereas LP/CP-based methods are more suitable when near-optimal accuracy is required and additional computation is affordable. Flow-based methods remain competitive for exact computation when effective structural pruning substantially reduces the flow network.

\subsubsection{Solutions for original DDS problem}

Tables \ref{tab:sum-exact-dds} and \ref{tab:sum-approx-dds} summarize exact and approximation algorithms for DDS.
DDS is more complex than UDS due to the ratio $c=|S|/|T|$ between two vertex sets, which has $O(n^2)$ possible values.

\begin{table}[h]
  \centering
  \caption{Summary of exact DDS algorithms, $k\leq n^2$, $h\leq n^2$, $t_{\textsf{Flow}}$ denotes the time cost of the max-flow algorithm.}
    \begin{tabular}{c|l|l}
    \hline
    Category & Algorithm & Time Complexity \\
    \hline
    \hline
    \multirow{2}[0]{*}{LP-based} & {\tt LP-Exact} \cite{charikar2000greedy}  & $\Omega(n^6)$ \\
    \cline{2-3}
          & {\tt CP-Exact} \cite{ma2022convex}  & $O(h\cdot t_{\textsf{Flow}})$ \\
    \hline
    \multirow{2}[0]{*}{Flow-based} & {\tt Flow-Exact} \cite{khuller2009finding}  & $O(n^2 \log n \cdot t_{\textsf{Flow}} )$ \\
    \cline{2-3}
          & {\tt DC-Exact} \cite{ma2020efficient}  & $O(k\log n \cdot t_{\textsf{Flow}})$ \\
    \hline
    \end{tabular}%

  \label{tab:sum-exact-dds}%
\end{table}%

\begin{table*}
  \centering
  \caption{Summary of approximation DDS algorithms, $\epsilon>0$, $\delta>1$, and $t_{\textsf{FW}}$ denotes the time cost of the Frank-Wolfe algorithm.}
    \begin{tabular}{c|l|c|l}
    \hline
    Category & Algorithm & Approx. ratio & Time complexity \\
    \hline
    \hline
    \multirow{3}[0]{*}{Peeling-based} 
          & {\tt BS-Approx} \cite{charikar2000greedy} & 2 & $O(n^2\cdot(n+m))$ \\
    \cline{2-4}
          & {\tt KS-Approx} \cite{khuller2009finding} & $\geq 2$  & $O(n+m)$ \\
    \cline{2-4}
          & {\tt PM-Approx} \cite{bahmani2012densest} & $2\delta(1+\epsilon)$  & $O(\log_{\delta}{n} \log_{1+\epsilon}{n} \cdot (n+m))$ \\
    \hline
    Core-based  & {\tt Core-Approx} \cite{ma2020efficient} & 2 & $O(\sqrt{m}\cdot(n+m))$ \\
    \hline
    LP-based  & {\tt CP-Approx} \cite{ma2022convex} & $1+\epsilon$ & $O(\log_{1+\epsilon}n\cdot t_{\textsf{FW}})$ \\
    \hline
    Flow-based  & {\tt Flow-Approx} \cite{chekuri2022densest} & $1+\epsilon$ & $\tilde{O}(\frac{m}{\epsilon^{2}})$ \\
    \hline
    \end{tabular}%

  \label{tab:sum-approx-dds}%
\end{table*}%

%

Exact algorithms enumerate all possible $c$ and solve a corresponding UDS instance. This leads to high complexity, e.g., $\Omega(n^6)$ for LP-based methods and $O(h \cdot t_{\textsf{Flow}})$ for flow-based ones.
To improve efficiency, {\tt DC-Exact}~\cite{ma2020efficient} reduces both the search space and flow network size using $[x,y]$-core decomposition and divide-and-conquer. 
{\tt CP-Exact}~\cite{ma2022convex} further reduces LP calls via relaxation and dual optimization with Frank-Wolfe.
Overall, exact methods focus on reducing the enumeration of $c$ and the size of the flow network.

Since the exact algorithms are still costly to handle large-scale directed graphs, so many efficient approximation algorithms of DDS problem have been developed, including peeling-based~\cite{charikar2000greedy,khuller2009finding,bahmani2012densest}, core-based~\cite{ma2020efficient}, LP-based~\cite{ma2022convex}, and flow-based~\cite{chekuri2022densest} algorithms.
The peeling-based methods obtain the approximate solution by greedily searching all possible subgraphs with the maximum density corresponding to $c$.
The core-based algorithm proves that the $[x, y]$-core with the largest $x \cdot y$ among all $[x, y]$-cores is a solution of DDS with an approximation of 2.
The LP-based algorithm is similar to {\tt LP-Exact}, but it is not necessary to use the maximum flow algorithm to obtain an exact solution.
The flow-based approximation algorithm obtains an approximate result by performing partial max-flow computations.

{\color{black}

The flow-based approximation algorithm obtains an approximate result by performing partial max-flow computations.

{\color{black}
}

\textbf{Practical comparison.} For DDS, practical performance depends not only on the cost of solving each fixed-ratio instance, but also critically on the number of source--sink ratios $c=|S|/|T|$ that must be examined. Empirical results in~\cite{zhou2024depth} show that core-based graph reduction and divide-and-conquer can substantially reduce both the candidate graph and the number of ratio values, leading to speedups of over two orders of magnitude on some large directed graphs. This explains why methods with similar asymptotic guarantees may exhibit very different practical performance. When moderate accuracy is sufficient, core-based methods generally provide the best scalability with a factor-$2$ guarantee. For higher accuracy, convex-programming-based methods provide $(1+\epsilon)$ guarantees, but smaller $\epsilon$ increases the number of iterations and ratio configurations that must be processed. Exact methods are practical mainly when structural pruning effectively reduces the ratio search space and flow-network size. Therefore, DDS algorithm selection should jointly consider accuracy, graph scale, and the effectiveness of core-based reduction rather than theoretical complexity alone.

\textbf{Selection guidelines.}
Across both UDS and DDS, algorithm selection should consider graph
scale, accuracy requirements, and computational resources rather than
theoretical complexity alone.
Peeling- or core-based methods are generally preferable for large-scale
graphs and latency-sensitive applications due to their low overhead.
When near-optimal accuracy is required, LP/CP-based methods provide a
better accuracy--efficiency trade-off, while exact flow-based methods
are attractive when the graph is moderate in size or can be
substantially reduced by pruning.
For iterative methods, smaller $\epsilon$ or a larger iteration budget
$T$ generally improves solution quality at the cost of additional
runtime.
For DDS, the number of candidate source--sink ratios introduces an
additional computational factor, making ratio pruning and core-based
reduction particularly important.





\subsection{Comparison of other variants of UDS}
In addition to variants for different graph types and densities, the UDS problem has some variants that can be classified into two groups with different constraints:

(1) Constraints on the number of output results: a) Density-friendly graph decomposition~\cite{tatti2015density, danisch2017large} enumerates all subgraphs in order of density, and DS deconstruction~\cite{chang2020deconstruct} enumerates all the DSs.
b) Finding at most $k$ subgraphs with certain conditions, such as top-$k$ locally DSs with maximum density~\cite{qin2015locally,ma2022finding}, the top-$k$ subgraphs with maximum total density~\cite{galbrun2016top,dondi2021novel,dondi2021top}, and at most $k$ subgraphs with maximum total density while limiting overlap between them~\cite{balalau2015finding}.

(2) Constraints on the size of the output results: These works include finding a DS of size $k$, a DS of size no less than $k$, and a DS of size no greater than $k$ \cite{asahiro2000greedily,bhaskara2010detecting,bhaskara2012polynomial,bourgeois2013exact,nonner2016ptas,kawase2018densest,gonzales2019densest,andersen2009finding}.

The first group of variants can be applied to applications that require the identification of multiple dense regions, such as community detection or predicting anomalous behavior in networks.
For instance, the DS deconstruction methods can be used to find the subgraphs with the highest density, which may reveal patterns or outliers in large graphs.
On the other hand, the second group of variants is suitable for applications that have specific requirements on the sizes of the results, such as event organization or identifying a fixed number of vertices for analysis, which can help with understanding the underlying structure of the graph or identifying regions of interest.

(3) Constraints on the connectivity of the output results.
Some recent studies (e.g., \cite{bonchi2021finding}) identify the DS with connectivity constraints, such as finding a $k$-vertex or $k$-edge connected subgraph with maximum edge density.
A $k$-vertex (or $k$-edge) connected subgraph remains connected after removing up to ($k-1$) vertices (or edges), ensuring high connectivity.

Overall, these variants with constraints on the output results provide a flexible and powerful set of tools for analyzing graphs in various real-world applications.
    \section{Related work}
\label{sec:related}

This section reviews related studies, including cohesive subgraph search and graph clustering.

\subsection{Cohesive subgraph search}
\label{sec:related-DSD}

{\color{black}
Cohesive subgraphs are closely related to DS and are widely used in
Web, financial, social, and biological networks.
Unlike DSD, which explicitly maximizes a density objective, cohesive
subgraphs are typically defined by structural constraints such as
minimum degree, triangle support, or edge connectivity.
Representative models include $k$-core
~\cite{batagelj2003m,seidman1983network}, $k$-truss
~\cite{cohen2008trusses,saito2008extracting,zhang2012extracting},
$k$-ECC~\cite{hu2016querying,yuan2017efficient}, and clique-based
variants~\cite{danisch2018listing,abello2002massive,
balasundaram2011clique,zhou2021improving}.
Among them, $k$-core is most closely related to DS: it constrains
minimum degree, whereas DS maximizes average degree, and the
$k_{\max}$-core provides a 2-approximation for UDS
~\cite{fang2019efficient}.

These cohesive models are fundamental to community search and were
systematically reviewed in the previous survey~\cite{fang2020survey}.
That survey primarily focuses on query-oriented subgraphs satisfying
prescribed cohesiveness constraints, whereas the present survey studies
density-maximization problems and their optimization algorithms.
Accordingly, this paper emphasizes density formulations, flow/min-cut,
LP/convex programming, peeling and approximation, dynamic maintenance,
and extensions to directed, weighted, hypergraph, and heterogeneous
settings.
Although cohesive structures can serve as approximations or pruning
structures for DSD, the two research lines differ fundamentally in
their objectives and algorithmic foundations.
}

Nevertheless, there is a lack of systematic reviews on DSD, except for a few preliminary works \cite{zhou2024depth,gionis2015dense,farago2019search,lanciano2023survey}.
The first two works \cite{gionis2015dense,farago2019search} briefly review the works in the general area of dense subgraph computation, with little attention on the topic of DSD.
The last one~\cite{zhou2024depth} is an experimental paper on UDS and DDS.
The third one~\cite{lanciano2023survey} is a survey of DSD in ACM Computing Surveys, but it differs from ours in three aspects:
(1) Our work covers more topics of DSD. For example, we have discussed topics like density-friendly decomposition and analyzing the relationships of different density definitions for different types of graphs, while \cite{lanciano2023survey} does not cover these topics.
(2) Unlike \cite{lanciano2023survey}, our work conducts a comprehensive comparison study in Section \ref{sec:comparison}, which offers a deeper understanding of the interrelationships among different works.
%
%
(3) Our work emphasizes practical guidance, unlike \cite{lanciano2023survey}, which is more theoretical. In practice, exact algorithms suit small to medium graphs, while approximation methods are preferred for large graphs due to better efficiency and scalability.

\subsection{Graph clustering}
Graph clustering is a well-studied problem in data mining. The problem aims to partition the graph into disjoint subsets. It is useful in many real-world applications such as marketing, customer segmentation, data summarization, and community detection.

A series of different methods have been proposed for identifying clusters, such as 
min-max cut methods~\cite{shi2000normalized},
hierarchy methods~\cite{girvan2002community}, 
structural-based methods~\cite{xu2007scan,ruan2021dynamic},
spectral-based methods~\cite{von2007tutorial,tang2009scalable}, modularity maximization-based methods~\cite{blondel2008fast}, random walks~\cite{pons2005computing}, graph partition~\cite{karypis1995metis}, embedding~\cite{perozzi2014deepwalk}, label propagation~\cite{gregory2010finding}, centrality~\cite{newman2004finding}, locality sensitive hashing~\cite{macropol2010scalable}, deep learning~\cite{yang2016modularity}, information diffusion~\cite{hajibagheri2012community} and other methods~\cite{henderson2010hcdf}. 
Most of these works use a global predefined criterion for generating subsets.
The detailed investigation of graph clustering in undirected graphs can refer to existing surveys and empirical evaluations~\cite{amelio2014overlapping,kim2015community,yang2012defining}.


\section{Future research directions}
\label{sec:future}

{\color{black}

In this section, we discuss several promising research directions that may further advance the study of DSD.

\noindent$\bullet$ \textbf{DSD on HINs.}
As summarized in Table \ref{tab:models_others}, the original DSD problems on undirected and directed graphs have been extended to several specialized graph models, including bipartite graphs, multilayer graphs, uncertain graphs, and dual graphs. Many of these models can be viewed as special cases of HINs, where vertices and edges may belong to multiple semantic types. Such heterogeneous structures are widely observed in knowledge graphs, bibliographic networks, and biological interaction networks.

A key challenge is to design density measures that can capture both structural connectivity and heterogeneous semantics. Existing studies such as \cite{chen2023densest} investigate DSD in HINs through meta-path based connectivity patterns \cite{sun2011pathsim}. However, these definitions are not fully consistent with density notions used in homogeneous graphs and often rely on manually designed meta-paths. A promising direction is therefore to explore more general density formulations that incorporate heterogeneous structural patterns, such as motif-based structures \cite{hu2019discovering}, typed subgraph patterns, or relational constraints \cite{jian2020effective}.

\textit{Open problems.} Important questions remain largely unexplored. For example, how to design density measures that remain comparable across different type configurations, how to control the combinatorial explosion of heterogeneous patterns, and how to develop scalable algorithms for generalized heterogeneous density models.

\noindent$\bullet$ \textbf{Efficient DSD algorithms.}
Despite significant progress in algorithm design, improving the scalability of DSD algorithms remains an important research challenge. Many exact algorithms rely on flow-based techniques, which can become computationally expensive on massive graphs. In addition, approximation algorithms often introduce parameters that influence both solution quality and running time, making it difficult to achieve a consistent balance between efficiency and accuracy in large-scale settings.

Future work may focus on several directions. First, parallel and distributed implementations of DSD algorithms can significantly improve scalability by leveraging modern multi-core and distributed computing platforms. While some progress has been made in parallel approximation algorithms \cite{bahmani2012densest}, the design of efficient parallel exact algorithms remains largely unexplored. Second, it is important to investigate algorithmic frameworks that reduce the dependence on repeated global computations, such as localized updates or incremental maintenance strategies, which may significantly improve performance in large and dynamically evolving graphs.

\textit{Open problems.} Key challenges include identifying the computational bottlenecks of current flow-based and peeling-based frameworks, designing algorithms with improved scalability while preserving approximation guarantees, and developing practical implementations that perform robustly on graphs with billions of edges.

\noindent$\bullet$ \textbf{Application-driven variants of DSD.}
Another promising direction is to develop DSD variants tailored to specific application scenarios. Although many variants of the DSD problem have been proposed, most are motivated by theoretical generalizations rather than application requirements. Incorporating domain-specific constraints into the density model may lead to more practical formulations.

For example, DSD has been widely used for detecting dense communities in networks \cite{chen2010dense}. In geo-social networks, however, communities often exhibit both structural connectivity and spatial proximity \cite{fang2017effective}. This suggests that density measures may need to incorporate spatial information, such as distance-based edge weights or locality constraints. Similar challenges also arise in other application domains, where temporal dynamics, resource constraints, or heterogeneous attributes may influence the definition of dense substructures.

\textit{Open problems.} An important research direction is to understand how application constraints, such as spatial proximity, temporal dynamics, or attribute similarity, can be systematically integrated into density formulations while maintaining algorithmic tractability.
}

\section{Conclusion}
\label{sec:conclusion}

In this paper, we conduct a comprehensive review of the topic of DSD on large graphs by reviewing around 50 research articles focusing on this topic between 1984 and 2023.
We first introduce the typical applications and key challenges of DSD.
We then classify existing works of DSD according to their definitions, and for each class of works, we systematically review and discuss the representative DSD solutions on undirected graphs, directed graphs, and other graphs, respectively.
We also discuss the representative variants of DSD problems and solutions over different kinds of graphs.
Finally, we point out a list of promising future research directions of DSD.
In summary, our survey provides an overview of the start-of-the-art research achievements on the topic of DSD, and it will give researchers a thorough understanding of DSD.

		

\bibliographystyle{IEEEtran}
\bibliography{densest}

\end{document}